\documentclass{optica-article}

\journal{opticajournal} 

\articletype{Research Article}

\usepackage{lineno}
\linenumbers 
\usepackage{xr}

\begin{document}
\title{Dynamically reconfigurable THz quantum walk comb laser through subharmonic excitation}
\author{Valerio Digiorgio\authormark{1}*, Robert M. Gray\authormark{1}, Marco Raffa\authormark{1}, Paolo Micheletti\authormark{1}, Alexander Dikopoltsev\authormark{1}, Mattias Beck\authormark{1}, J\'er\^ome Faist\authormark{1}, Giacomo Scalari\authormark{1}*}

\address{\authormark{1}Institute for Quantum Electronics, Department of Physics, ETH Z\"urich, Z\"urich, Switzerland\\
}

\email{\authormark{*}vdigiorgio@ethz.ch, gscalari@ethz.ch} 


\begin{abstract*} 

On-chip frequency combs are increasingly relevant to both laser science and applications. Broad bandwidths and flat-top spectral envelopes are especially desirable for precision spectroscopy and dense wavelength-division multiplexed communications. Toward these goals, active microwave modulation has emerged as a powerful strategy for generating, stabilizing, and reconfiguring frequency combs at the source. However, practical challenges associated with high-frequency modulation imposes an upper bound on the accessible cavity free spectral ranges.\
Here, we demonstrate a subharmonic locking scheme in a quantum walk comb laser, a recently introduced platform for broadband and highly controllable comb states. Using a THz ring quantum cascade laser, we realize quantum walk comb formation under strong microwave injection at successive subharmonics of the cavity round-trip frequency, tuning the comb spacing from 15.8 to 1.58 GHz. The resulting states arise from fast-gain dynamics and nonlinear microwave mixing in the laser cavity. Through two-tone injection, we exploit this mixing to dynamically control the comb bandwidth and spectral shape. These results establish subharmonic excitation as a route to broadband, reconfigurable semiconductor comb generation in high-FSR cavities.

\end{abstract*}

\section{Introduction}

Optical frequency combs have demonstrated benefit in many application areas, ranging from metrology~\cite{hansch2006nobel} and precision spectroscopy~\cite{picque2019frequency,coddington_dual-comb_2016} to high-speed communications~\cite{Okawachi2023_rev}, among others~\cite{Kipp_Diddams_Holz_Science2011}. Broad deployment of frequency comb technologies demands compact and stable comb sources, which has driven substantial effort into their realization on-chip~\cite{chang2022integrated,scalari_-chip_2019}. Examples of integrated comb sources include electro-optic combs~\cite{parriaux2020electro, zhang2019broadband}, soliton microcombs~\cite{kippenberg_dissipative_2018, englebert2026temporal, herr_temporal_2014}, quantum cascade lasers \cite{Faist:2016eg, hugi_mid-infrared_2012, burghoff_terahertz_2014} and integrated mode-locked lasers~\cite{qiu_mamyshev_Kip_2026, guo2023ultrafast,davenport2018integrated}. \

To meet various application needs, flexible control over the comb properties, including the repetition rate, bandwidth, and spectral shape, is desirable. On the one hand, such control can be achieved in two stages by first generating a broadband frequency comb and subsequently performing spectral shaping~\cite{ferdous2011spectral,liu2025ultracompact}. However, this increases system size and complexity and typically reduces overall efficiency through the use of spectral filtering. Efforts to overcome these tradeoffs have leveraged active mechanisms for comb generation to enable spectral control at the source, including spectral shaping through the harmonic modulation of electro-optic combs~\cite{song2026universal} and continuous tuning of the comb tooth spacing via strong modulation in quantum cascade laser (QCL) devices~\cite{Jaidl:2025:tunable,senica2026continuously}.

Within this framework, the quantum walk comb laser (QWCL) has emerged as a particularly promising platform.\
The QWCL has recently been demonstrated as a novel scheme for active comb generation, exploiting the interplay between a phase modulation at the cavity free spectral range (FSR), or harmonics thereof, and a fast laser gain~\cite{heckelmann2023quantum}. Formation of the QWC requires that the free-running laser operate in a single mode, with subsequent modulation inducing coupling of the cavity modes to yield a frequency comb, the dynamics of which resemble a quantum walk on the synthetic lattice of frequency modes~\cite{heckelmann2023quantum,Dikopoltsev2025theoryQW}. Defect-free ring QCLs are the ideal candidates to exploit this technique, with the first such demonstration being realized in a mid-infrared QCL~\cite{heckelmann2023quantum}. Their active medium provides the ultrafast gain dynamics required for QWC formation, while the suppression of backscattering enables spontaneous symmetry breaking between the two degenerate counter-propagating modes, naturally leading to single-mode free-running operation.

This class of devices has attracted growing attention due to its high stability and controllability as an on-chip integrated frequency comb source, operating in both fundamental and harmonic regimes~\cite{marzban2026quantum,piciocchi2026spectral}. Recent efforts include demonstrations of high-power operation through improved outcoupler design~\cite{cargioli2025quantum,letsou2026high} as well as QWC formation in interband semiconductor lasers, traditionally considered to have slow gain recovery, enabling operation at telecommunications wavelengths~\cite{marzban2026quantum}. Moreover, flexible control over the comb has been realized through harmonic driving~\cite{marzban2026quantum} as well as spectral envelope shaping through two-tone harmonic modulation, which provides an additional degree of control and flexibility~\cite{piciocchi2026spectral}.
However, given the high FSRs of typical integrated devices, such harmonic driving places stringent demands on the surrounding radio frequency (RF) infrastructure. By contrast, subharmonic injection locking enables oscillator synchronization using a drive below the fundamental oscillation frequency. It has been extensively studied in electronic oscillators and phase-locked loops\cite{Lee2009subharmoniclockingPLLs}, where a subharmonic drive ultimately stabilizes the oscillator at its fundamental frequency. Related synchronization phenomena have also been reported in both amplitude modulated (AM) and frequency modulated (FM) optical systems such as Q-switched fiber lasers\cite{Wu2000rationalinjection,Chang2011fibersubharmonic}, mode-locked semiconductor lasers\cite{Hoshida1996subharmonicpulsetrain,Hoshida1996SH-ML,Teshima1998subharmonicMLLD}, and more recently in THz quantum cascade lasers\cite{wu2024harmonicsubharmonic,liu2025farey}.

\begin{figure}[!b]
	\centering
	\includegraphics[width=0.9\linewidth]{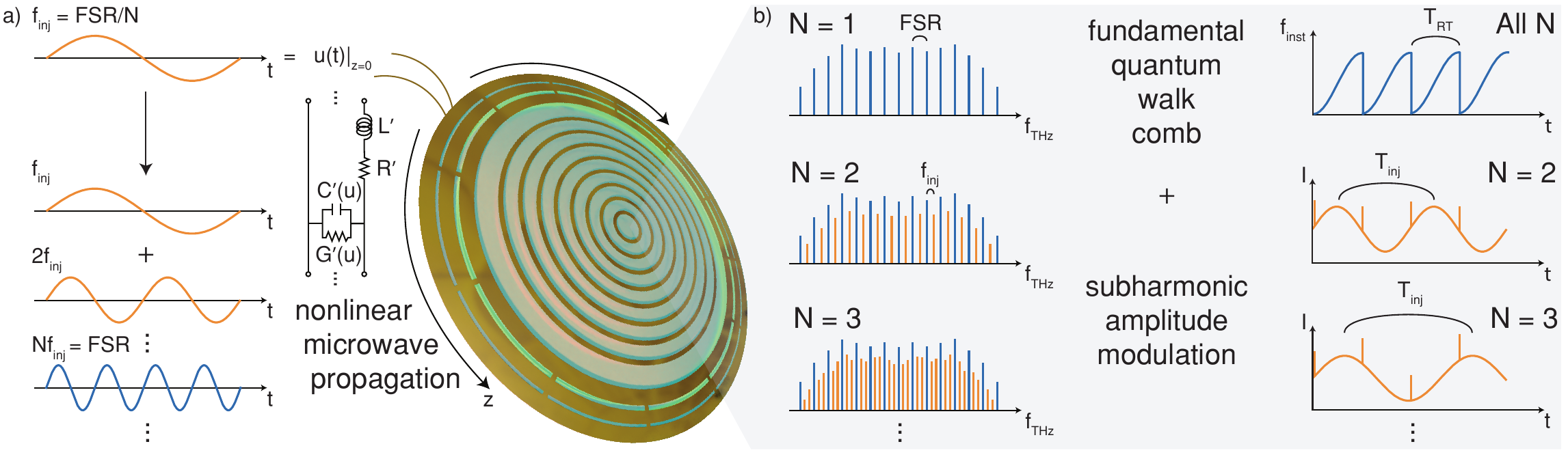}
	\caption[Quantum walk comb formation under subharmonic injection]{Quantum walk comb formation under subharmonic injection. (a) The THz quantum cascade laser, consisting of a double-metal ring resonator, is modulated at a frequency $f_\text{inj} = \text{FSR}/N$, the $N^\text{th}$ subharmonic of the THz free spectral range. Microwave nonlinearities arising due to the dependence of the distributed conductance, $G'(u)$, and capacitance, $C'(u)$, of the active region on the microwave voltage, $u(z,t)$, lead to upconversion of the injected tone. (b) For fundamental injection, with $f_\text{inj} = \text{FSR}$, a THz QWCL is realized. With subharmonic driving, the $N^\text{th}$ upconverted microwave at $Nf_\text{inj} = \text{FSR}$ similarly drives QWC formation. Thus, the QWCL under subharmonic injection maintains the characteristic half-cosine chirp with a period $T_\text{RT} = 1/\text{FSR}$. However, subharmonic injection also facilitates amplitude modulation at a period $T_\text{inj} = 1/f_\text{inj}$, leading to the proliferation of additional comb teeth with spacing equal to the injection frequency. $L'$, distributed inductance; $R'$, distributed resistance.}
	\label{fig:SubQW comb_Concept}
\end{figure} 

In this work, we combine the broadband and highly controllable QWC regime with subharmonic RF driving. We demonstrate QWCL operation under excitation at successive subharmonics of the cavity FSR, from the fundamental round-trip frequency of 15.8 GHz down to 1.58 GHz, where the comb repetition rate is directly stabilized at the externally imposed subharmonic frequency. The coherence of the observed subharmonic states is assessed via shifted-wave interference Fourier transform spectroscopy (SWIFTS) measurements~\cite{burghoff_terahertz_2014}, which reveal that the subharmonic states retain the characteristic frequency-modulated evolution of the fundamental QWC, while acquiring an amplitude-modulation contribution at the period of the injected subharmonic. We implement this scheme in the THz spectral range using a ring QCL based on a planarized double-metal waveguide~\cite{senica2022planarized}. A passive bullseye antenna integrated with the laser provides mW-level output powers despite the absence of scattering centers~\cite{micheletti2023terahertz}, enabling stable QWCL operation with broad, flat-top spectra around a carrier frequency of 3 THz.

To understand our experimental results, we extend the previously established QWCL theory~\cite{Dikopoltsev2025theoryQW,opacak_theory_2019,burghoff_unraveling_2020} along with the transmission-line approach used in ref.~\cite{Schreiber2025_model} for accurate modeling of the injected microwaves by considering the laser as a nonlinear transmission line, in which the bias-dependent electrical response of the quantum-cascade active region generates additional frequency components through harmonic generation and microwave mixing (see Fig.\ref{fig:SubQW comb_Concept}). Under excitation at a subharmonic of the cavity FSR, the generated harmonic resonant with the FSR drives QWC formation, while the injected subharmonic directly sets the comb mode spacing. Through two-tone injection, we exploit this nonlinear microwave mixing to dynamically control the comb bandwidth and spectral distribution, realizing unprecedented spectral control in a monolithic semiconductor laser source. This is also advantageous for photonic integration of frequency comb sources \cite{chang2022integrated}, as the same comb features can be realized in much shorter cavities with larger FSR, leading to a reduced device footprint and lower power consumption, without necessitating an extremely high-frequency RF source.

\section{Results}
\subsection{The THz quantum walk comb laser}
We employ an 800 $\mu$m radius double-waveguide ring QCL featuring a bullseye antenna, similar to the devices reported in \cite{micheletti2023terahertz}. In contrast to the previous study, this laser emits as a single mode throughout its entire operating range and does not exhibit a hysteretic behavior, i.e. multi-mode free-running operation cannot be induced by fast variations of the bias current or cycling the RF injection power. This behavior can be ascribed to the absence of defects, making this device a good candidate to behave as a QWCL.

\begin{figure}[!htb]
	\centering
	\includegraphics[width=0.9\linewidth]{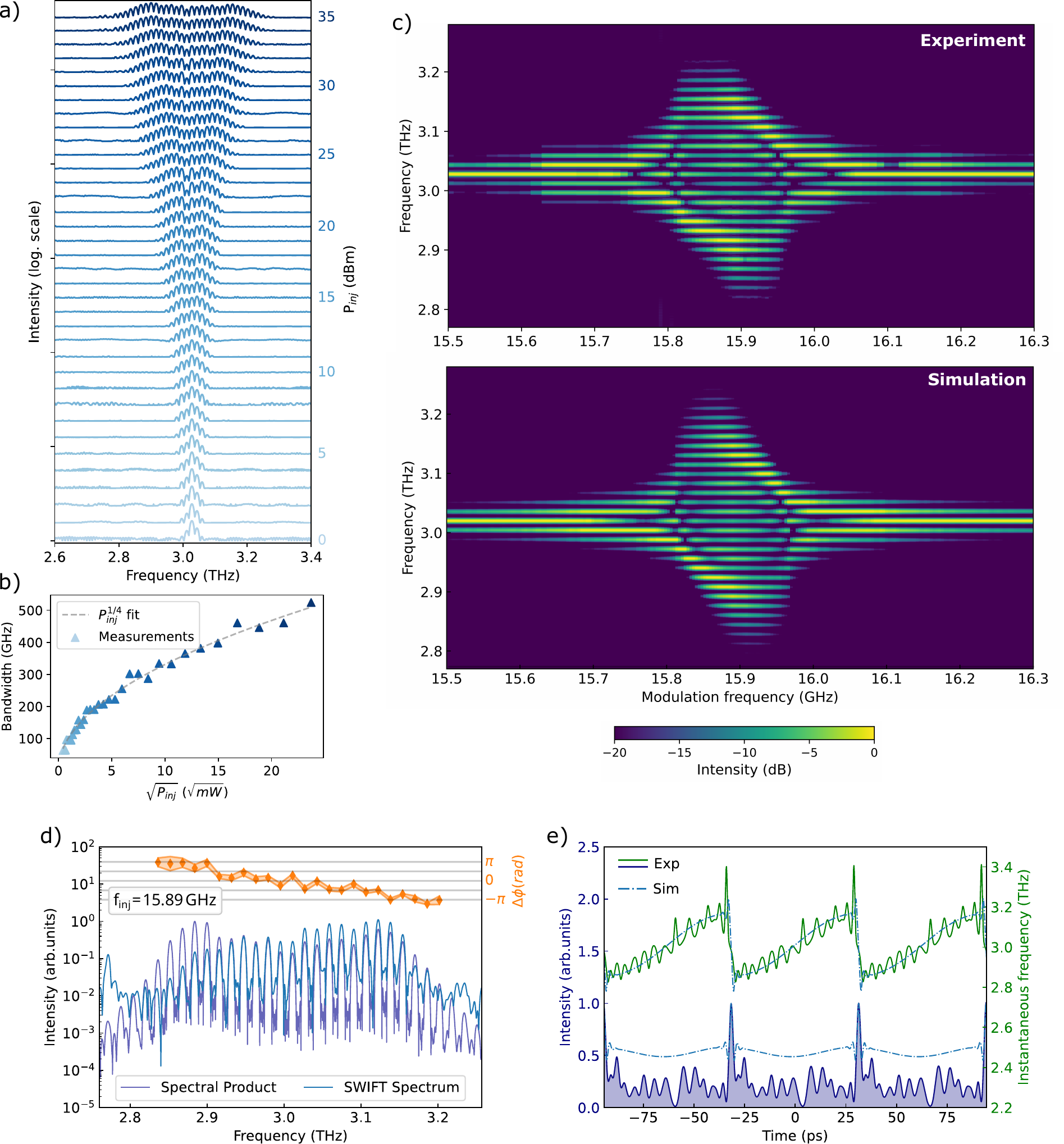}
	\caption[Analysis of single mode ring QCL under strong RF injection]{QWCL operation in an 800 $\mu$m radius, coupled-waveguide ring QCL under different modulation conditions. The laser is biased at 9.5 V and operated at 40K. (a) Device spectra collected at an injection frequency of 15.89 GHz; the injected power is swept between 0 dBm and 35 dBm. The measured bandwidth of each spectrum is reported in (b, triangles) as a function of the square root of the injected power. The data are fitted with a $P^{1/4}$ (grey dashed line) dependence. (c) Experimental and simulated spectra of the device operating in the same conditions for different RF injection frequencies under strong modulation (35 dBm). (d) SWIFT spectroscopy measurement showing the overlap between spectral product and SWIFT spectrum together with the intermodal phase difference. (e) Temporal reconstruction of the laser output. The measured data (solid line) and simulated trace (dash-dotted line) show the characteristic QWCL features, with a half-cosine modulation of the instantaneous frequency.}
	\label{fig:QW_expandsim}
\end{figure}

To investigate the QWC behavior, we inject the ring QCL at its FSR, yielding $\sim$ 400-GHz broad spectra around the center frequency of 3 THz. It further features an extremely flat envelope, in good agreement with what was observed in mid-IR. We also see a progressive broadening of the spectra with increasing injection power, following the expected dependence on the fourth root of the injected power, as predicted by QWCL theory (see Fig.~\ref{fig:QW_expandsim}a-b)~\cite{Dikopoltsev2025theoryQW}.

In Fig.~\ref{fig:QW_expandsim}c, we show the output spectrum as a function of detuning for both the experiment (top) and corresponding simulation (bottom, see Supplementary Information for simulation details), exhibiting the characteristic expansion as the resonant frequency is approached. Notably, the injection map appears slightly asymmetric, as the center of the spectra on the low frequency side of the plot has a lower mean frequency than the ones at positive $f_{\text{inj}} - f_{\text{rep}}$. This is attributed to dispersion, which shifts the cavity modes across the spectrum, affecting the RF-induced mode coupling. For the measured device, which exhibits a positive GVD, injection at $f_{\text{inj}} < f_{\text{rep}}$ preferentially locks the higher frequency modes which feature a smaller cold cavity FSR compared to the lower frequency ones.

To study the dependence of the inter-modal phase profile on the injection frequency, we performed SWIFT spectroscopy measurements with a fast Schottky detector (as in Ref.\cite{senica2026continuously}) at various injection frequencies, ranging from off-resonant to resonant injection (see Supplementary Information). The distribution of the inter-modal phases is well reproduced by simulations, showing the theoretically predicted half-cosine chirp of the instantaneous frequency when driving at resonance (Fig.~\ref{fig:QW_expandsim}). This profile continuously changes by detuning the injection from the cavity round-trip frequency, reaching a continuous sinusoidal phase modulation in the highly off-resonant case, as predicted by the QWCL theory. We will refer to the on-resonance frequency comb as the `fundamental comb' in the following.
The discrepancies between the spectral product and the SWIFT spectrum in Fig.~\ref{fig:QW_expandsim}e are not attributed to a lack of coherence but rather to the diffractive nature of the bullseye antenna emission. This was checked by performing SWIFT spectroscopy measurements with different alignment conditions, which resulted in a different spectral collection efficiency. The peak in the SWIFT spectrum below 2.8 THz originates from noise and is not considered for the time reconstruction of the instantaneous frequency.

\begin{figure}[!b]
	\centering
	\includegraphics[width=\textwidth]{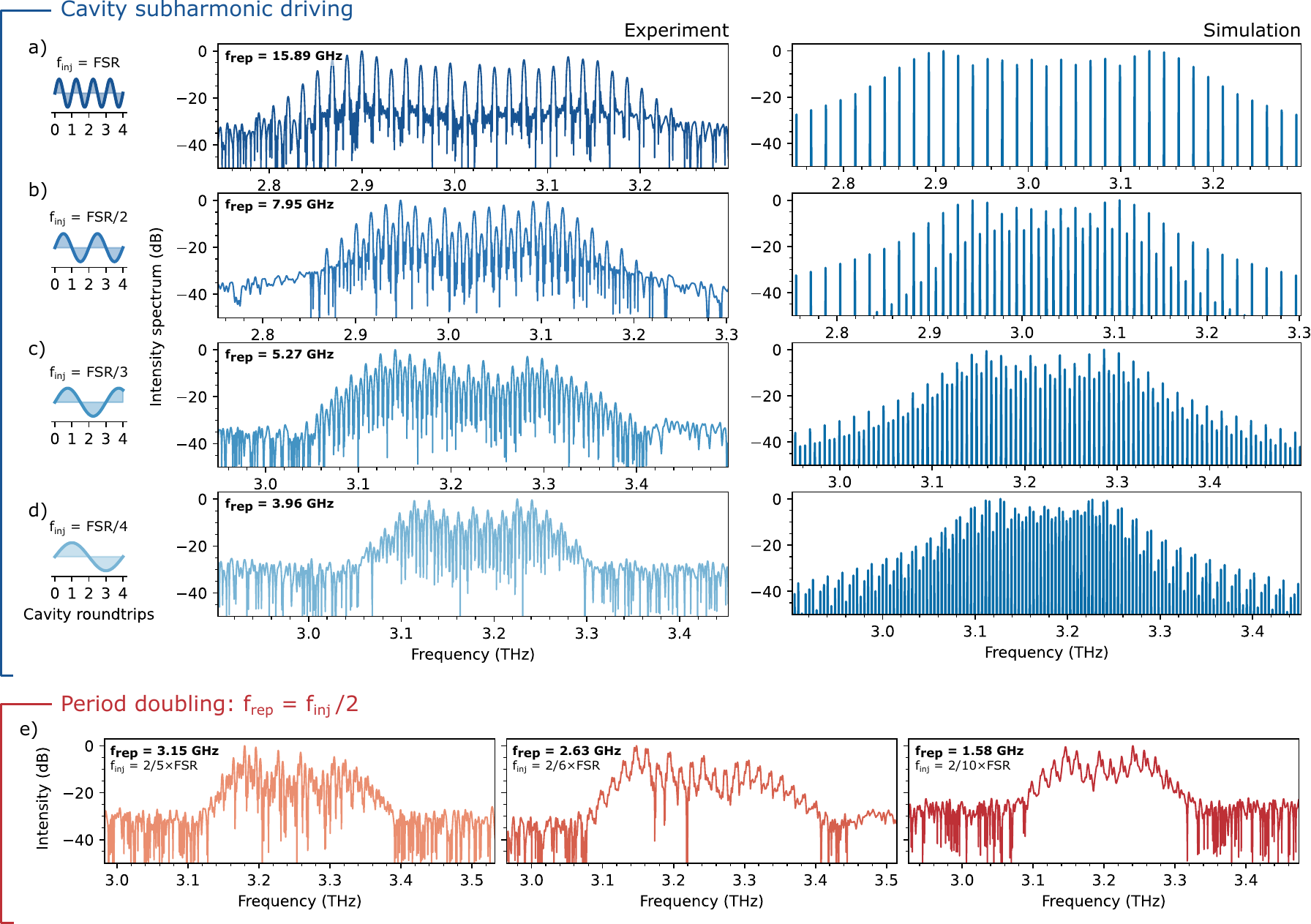}
	\caption[]{Sub-FSR modulation regime results. The QCL is operated at 40 K at different bias voltages ranging from 9.5 V to 10.1 V. The injected RF power ranges from +30 dBm to +35 dBm, and the injected frequency is indicated, together with the mode separation, in each plot. The device locks at the injected frequency (upper panel) or at its half (lower panel), and always at a subharmonic of the round-trip frequency. (a-d) Cavity subharmonic microwave driving. Illustration of the modulation evolution (left) matching the injection frequency of the measured subharmonic frequency comb spectra (center), respectively for the fundamental state and the $1/N$ subharmonic states, with $N=2,3,4$. Simulated comb spectra for the same modulation conditions (right). (e) Period doubling dynamics, with mode spacing equal to half the injected frequency. Measured comb spectra with repetition rate matching FSR/5 (left), FSR/6 (center), and FSR/10 (right). Subharmonic states with $n>6$ have a mode spacing beyond the resolution of the FTIR used for the characterization. We just show here the spectral envelope for the $n=10$ case, without resolving the different lines. All spectra, measured and simulated, are plotted over a 550 GHz frequency span, with different central frequency.}
	\label{fig:Subharmonic_spc}
\end{figure}

\subsection{Subharmonic RF injection}

Significant spectral broadening is also observed under strong microwave modulation ($>30$ dBm) at subharmonics $f_{\mathrm{RT}}/N$ of the round-trip frequency. Such a mechanism has not yet been reported in mid-IR ring QCLs. This difference can be attributed to the different architectures for THz and mid-IR devices, where the mode confinement is achieved with distinct strategies. Optical confinement in mid-IR QLCs typically relies on dielectric waveguides, where the refractive index contrast between the active region and the surrounding semiconductor cladding layers provides a good spatial overlap between the optical mode and the gain medium. Extending this approach to THz wavelengths would require prohibitively thick layers. Metallic waveguide confinement is the typical solution adopted for THz QCLs, most commonly in a double-metal waveguide geometry. In this configuration, the active region is sandwiched between two metals, similar to a microstrip transmission line, with no cut-off frequency for the fundamental TM$_{00}$ mode, and supports efficient microwave propagation along the cavity. In mid-IR ring QCLs, RF modulation is typically applied only to a section of the ring and cannot propagate effectively along the cavity, so only a fraction of the gain is efficiently modulated.

The subharmonic broadening we observe arises from the combination of this distributed microwave propagation with the intrinsic nonlinear electrical response of the quantum-cascade active region. This response contains both conductive and capacitive contributions: the conductive nonlinearity originates from the bias-dependent alignment of the electronic states and the photon-driven transport \cite{Forrer:2020dy}, while the capacitive contribution arises from the voltage-induced displacement of the carrier distribution relative to the doped region \cite{Hinkov:16}. As a result, the THz QCL behaves as a nonlinear microwave transmission line, in which a strongly injected subharmonic tone can generate additional electrical components resulting from cascaded harmonic generation and microwave mixing. When injecting at subharmonics of $f_{\mathrm{RT}}$, this mechanism eventually leads to the generation of an RF component resonant to the cavity FSR, which accounts for the observed spectral broadening (see Fig.\ref{fig:SubQW comb_Concept}).

Operating the device under subharmonic RF modulation produces flat-top broadband spectra with variable mode separation from the fundamental round-trip frequency $f_{\text{RT}}\approx15.8$ down to $f_{\text{RT}}/10\approx1.58$ GHz. In most cases, the comb spacing follows the injected frequency $f_{\text{inj}}$, consistent with the nonlinear microwave propagation picture described above: the bias-dependent electrical response of the active region generates harmonic components of the injected tone, allowing a component resonant with $f_{\text{RT}}$ to drive the QWCL while the subharmonic modulation sets the mode spacing (Fig. 3a-d). However, under selected operating conditions, we observe a distinct locking regime in which the comb repetition rate is further reduced to $f_{\text{rep}}=f_{\text{inj}}/2$ (Fig. 3e). This behavior is reminiscent of period-doubling dynamics, a mechanism which is expected in strongly nonlinear driven systems and has been observed in electrical oscillators as well as laser systems. However, the present nonlinear transmission line model does not reproduce stable locking at $f_{\text{inj}}/2$, although it captures the subharmonic states locked directly to $f_{\text{inj}}$. We therefore interpret the $f_{\text{inj}}/2$ states as evidence of an additional nonlinear dynamical instability of the electrically driven QCL, beyond the mechanisms included in the current simulations.

\begin{figure}[!htb]
	\centering
	\includegraphics[width=1\linewidth]{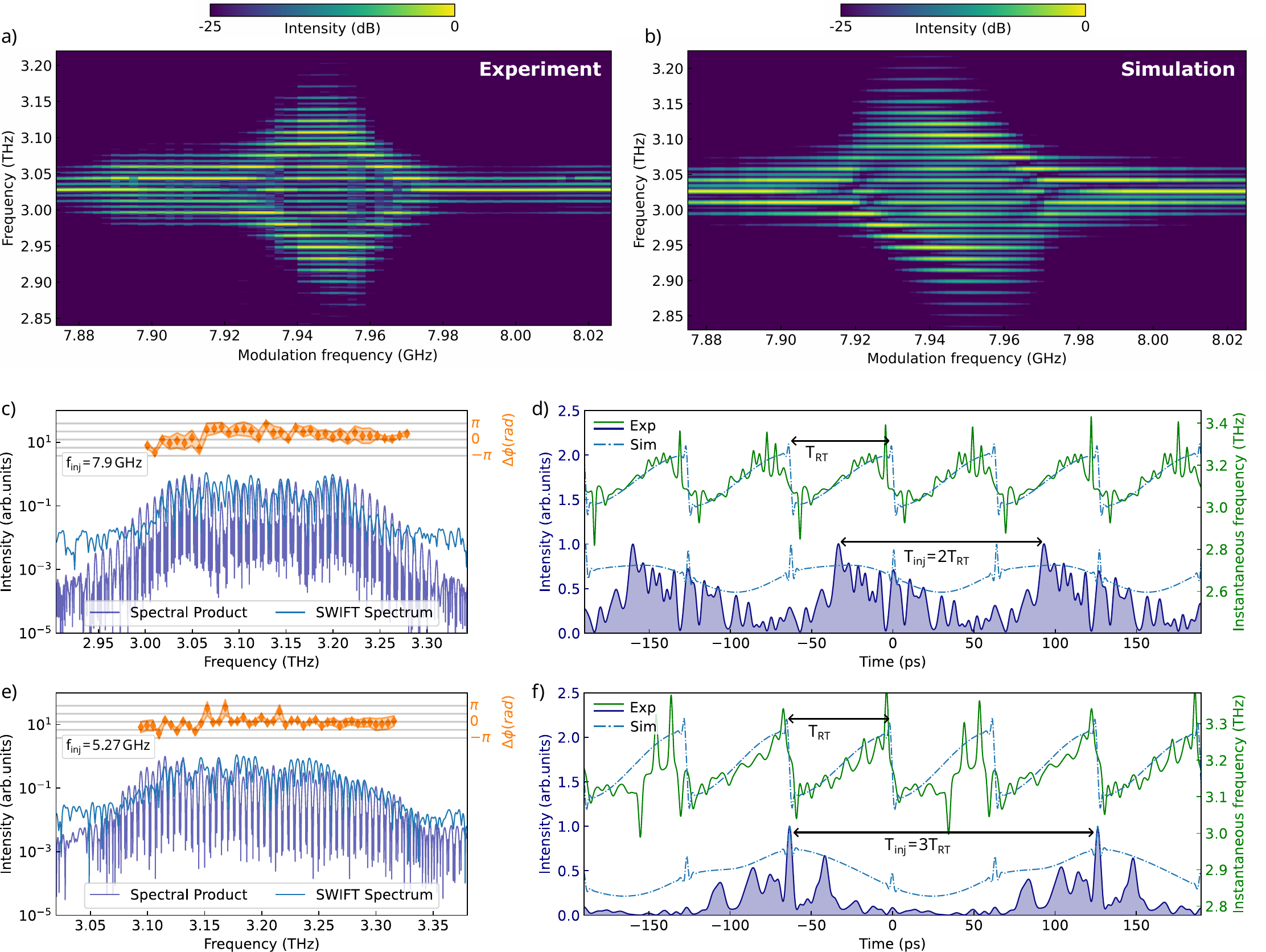}
	\caption[]{Subharmonic QWCL characterization. Measurement (a) and simulation (b) of the optical spectra of the ring QCL for different RF injection frequencies around the first subharmonic of the cavity FSR. The device is biased at 9.5V under strong modulation (+35 dBm). SWIFT spectroscopy measurements of the $1/2$ (c-d) and the $1/3$ (e-f) subharmonic state of the THz ring QCL. The device is operated at 40 K under strong microwave modulation (+30 dBm). The injected RF frequencies are indicated. The time reconstructions still show the characteristic QWCL evolution of the instantaneous frequency, with a full chirp repeating at every round trip, but the intensity follows the driving external microwave modulation, as predicted by the simulations.}
	\label{fig:AM-FM_QW}
\end{figure}

The measured spectra broaden significantly when the injected microwave tone is tuned close to the subharmonics of $f_{\text{RT}}$, in a similar way to the resonant spectral expansion shown in Fig. \ref{fig:QW_expandsim}. This is confirmed by the spectral maps measured as a function of the injection frequency (see Supplementary Information Fig.~S4). The specific case for the 1/2 subharmonic state is displayed in Fig.~\ref{fig:AM-FM_QW}a-b, where we show a qualitatively good agreement with the simulation.

The coherence of these subharmonic states is confirmed by the SWIFTS measurements, showing a good overlap between the spectral product and the SWIFT spectrum (see Supplementary Information Fig.~S7). This is especially true when investigating states with narrower bandwidth, but this is attributed to the spectral response of the bullseye antenna, as was found in the case of resonant modulation. The results of the SWIFTS analysis show a significant amplitude modulation contribution from the subharmonic modulation on top of the `fundamental' comb, which retains the frequency-modulated QWCL features. This behaviour is highlighted in Fig.~\ref{fig:AM-FM_QW}c-d for the 1/2 and 1/3 subharmonic states by the SWIFTS time reconstruction of the electric field intensity and instantaneous frequency. As for the fundamental QWC measurement in Fig.~\ref{fig:QW_expandsim}, the simulation of the intracavity field reproduces well the instantaneous frequency and exhibits qualitatively good agreement with the field intensity, exhibiting amplitude modulation at the period of the injected subharmonic.

\subsection{Two-tone RF injection}
\begin{figure}[!h]
	\centering
	\includegraphics[width=1\linewidth]{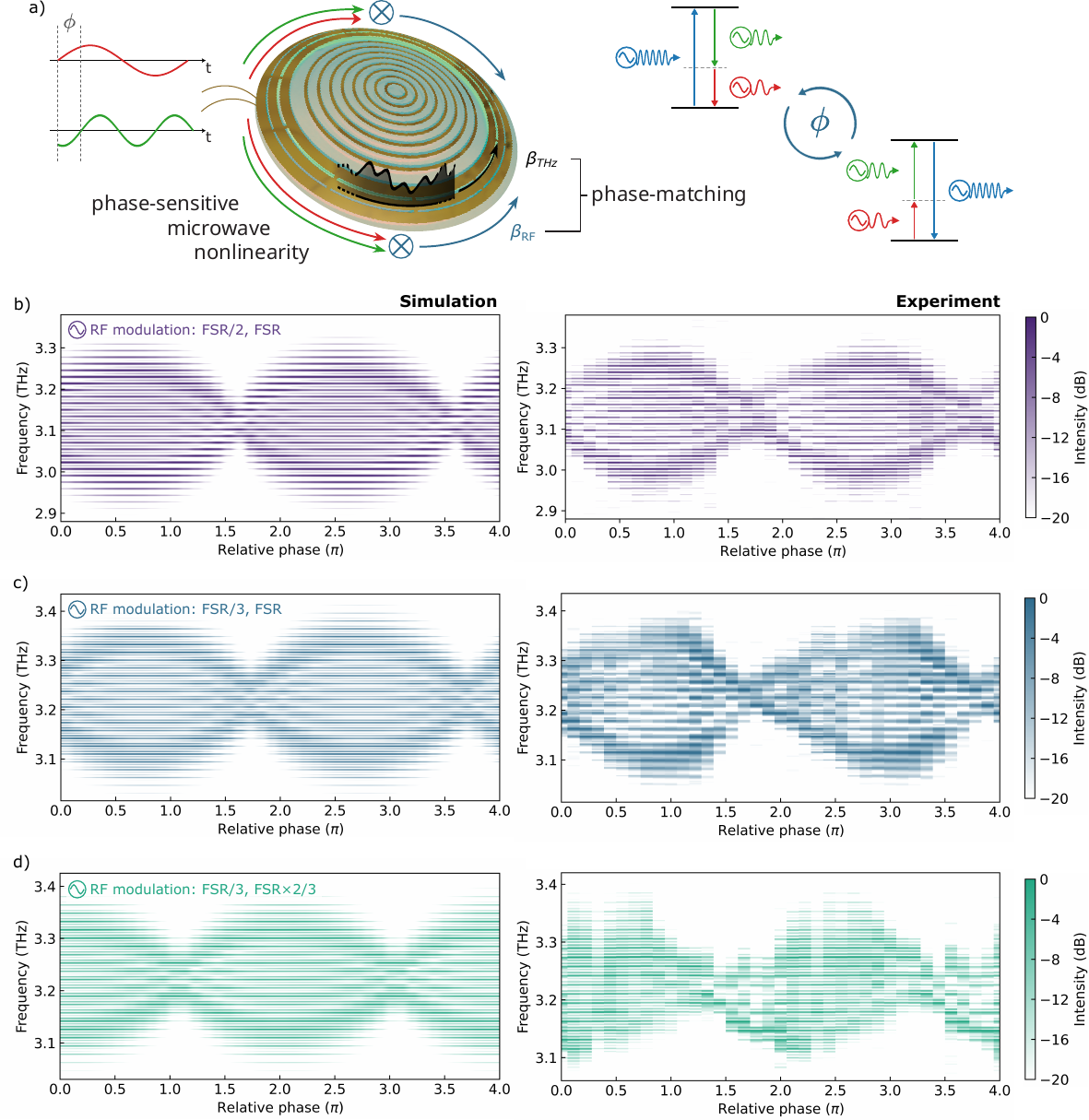}
	\caption[]{(a) Illustration of the phase-sensitive microwave nonlinear generation in the device. (Left) If two signals (red, green) are injected in the device, the nonlinearity will mix them generating new frequency components (blue). These will affect differently the traveling THz field (black) depending on the phase-matching between the two waves. (Right) Three-wave mixing processes are expected to occur in the device due to its electrical nonlinearity at microwave frequencies. In particular, the relative phases between the injected signals will affect the strength of the different microwave components. We expect a significant broadening of the laser spectrum when building up a strong electrical modulation resonant to the cavity FSR and satisfying the optimal phase-matching condition with the optical field. (b-d) Simulation (left) and measurements (right ) of the spectral shaping via injection of two RF tones through nonlinear mixing. (b) The QCL is operated at 40 K at 9.5 V under strong modulation (+35 dBm) with two microwave signals at FSR/2=7.9 GHz and FSR=15.8 GHz. (c) The laser is biased at 10.2 V and modulated at FSR/3=5.27 GHz and FSR=15.81 GHz. (d) The laser is biased at 9.5 V and modulated at FSR/3=5.27 GHz and FSR$\times$2/3=10.54 GHz.}
	\label{fig:Two-tone}
\end{figure}

Building on this nonlinear response, we further investigate the device under two-tone RF injection. While two-tone modulation at $f_{\mathrm{RT}}$ and $2f_{\mathrm{RT}}$ has recently been used to achieve spectral shaping in ring QCLs through control of the relative amplitudes and phases of the applied drives~\cite{piciocchi2026spectral}, the nonlinear electrical response of the THz QCL enables a distinct mechanism in the present device.

As discussed above, microwave propagation through the active region generates additional frequency components through nonlinear mixing. Under two-tone injection, up- and down-conversion processes can therefore generate a microwave component resonant with the cavity round-trip frequency, even when neither of the injected tones directly matches $f_{\mathrm{RT}}$. The net modulation experienced by the THz field depends on both the amplitudes and the relative phase of the injected signals, resulting in a corresponding control over the bandwidth and spectral distribution of the generated comb. Here, the net modulation is determined by the phase-matching condition between the microwave modulation and the THz field propagating in the ring. Therefore, only the microwave components whose phase velocity is matched to the group velocity of the THz field contribute efficiently to the round-trip phase modulation, as illustrated in Fig.~\ref{fig:Two-tone}a. Nonlinear microwave propagation can modify both the relative strength of the generated frequency components and their effective propagation constants, thereby changing how well this phase-matching condition is satisfied. In the two-tone regime investigated here, we find the latter effect to be the dominant mechanism underlying the observed comb shaping.

The availability of subharmonic driving considerably expands the range of frequency combinations that can participate in this nonlinear generation process, with representative examples shown in Fig.~\ref{fig:Two-tone}.The relative phase difference and the relative strength of the two modulations are the control knobs of this technique. The spectrum evolution is periodic, with a significant modification of the bandwidth. This effect is drastically suppressed when the two modulation amplitudes are greatly unbalanced.

\section{Conclusion}
We presented a ring QCL device exhibiting single-mode operation under free-running conditions, providing a suitable platform for exploring quantum-walk phenomena. The device exhibits the resonant behavior under strong RF injection typical of QWCLs, well reproduced by numerical simulations. In addition, subharmonic modulation of the investigated device leads to a new regime of rational subharmonic locking in mixed amplitude and frequency modulated comb states, arising from the interplay of intracavity microwave nonlinearities with the fast-gain laser dynamics. At the same time, it enables opportunities for two-tone modulation to realize advanced spectral shaping. Combining the techniques presented in this work, we reach an unprecedented level of control on the FSR, bandwidth, and peak-power frequency of the optical comb generated by a monolithic semiconductor source.

This injection locking scheme introduces several possibilities for this class of devices, both in terms of photonic integration and simplification of RF architecture. Specifically, by leveraging subharmonic driving, combs from compact and efficient high-FSR devices may be realized without necessitating a high-frequency microwave synthesizer. Key to maximizing the benefits will be engineering the nonlinear transmission line to support low-loss high-frequency microwave modes. Similarly careful engineering of the RF device properties could also enable extension of this technique to other wavelength ranges~\cite{calvar2013high}.

In conclusion, our results offer new insights on the QWCL dynamics in regimes beyond those previously accessed with mid-IR QCLs, while motivating further exploration of RF-mediated techniques for efficient frequency comb generation and control.

\begin{backmatter}
\bmsection{Funding}
The Authors acknowledge the use of FIRST clean room for laser fabrication, the SNF Project 200021-232335 for funding, as well as the EU Project iFLOWS. This work was partially performed within the framework of the 23FUN04 COMOMET project, supported by the European Partnership on Metrology, co-financed by the European Union’s Horizon Europe Research and Innovation Programme and by the Participating States. Funder ID 10.13039/100019599.

\bmsection{Acknowledgment}
V.D. and G.S conceived the idea. V.D. performed the measurements and data analysis with assistance from M.R. and P.M. R.M.G. developed the models and performed numerical simulations, with input from A.D and V.D. V.D., R.M.G. wrote the manuscript with support from G.S and A.D. V.D. and R.M.G. wrote the Supplementary Information. P.M. designed and fabricated the device. M.B. performed the epitaxial growth. G.S. and J.F. supervised the project and acquired funding.

\bmsection{Disclosures}
The authors declare no conflicts of interest.

\bmsection{Data availability} Data underlying the results presented in this paper are not publicly available at this time but may be obtained from the authors upon reasonable request.

\bmsection{Supplemental document}
See Supplement 1 for supporting content.

\end{backmatter}



\bibliography{QWcombTHz_ref}






\end{document}



\baselineskip24pt


\maketitle 

\section{Materials and Methods}

\subsection{Fabrication and design}
The laser chips were fabricated from MBE-grown wafers of the active material, consisting of a strongly diagonal, low-threshold broadband GaAs/AlGaAs heterostructure. The fabrication process follows a planarized waveguide technique, described in detail in Ref.\cite{senica2022planarized}.

\begin{figure}[t!]
\centering
\includegraphics[width=0.4\linewidth]{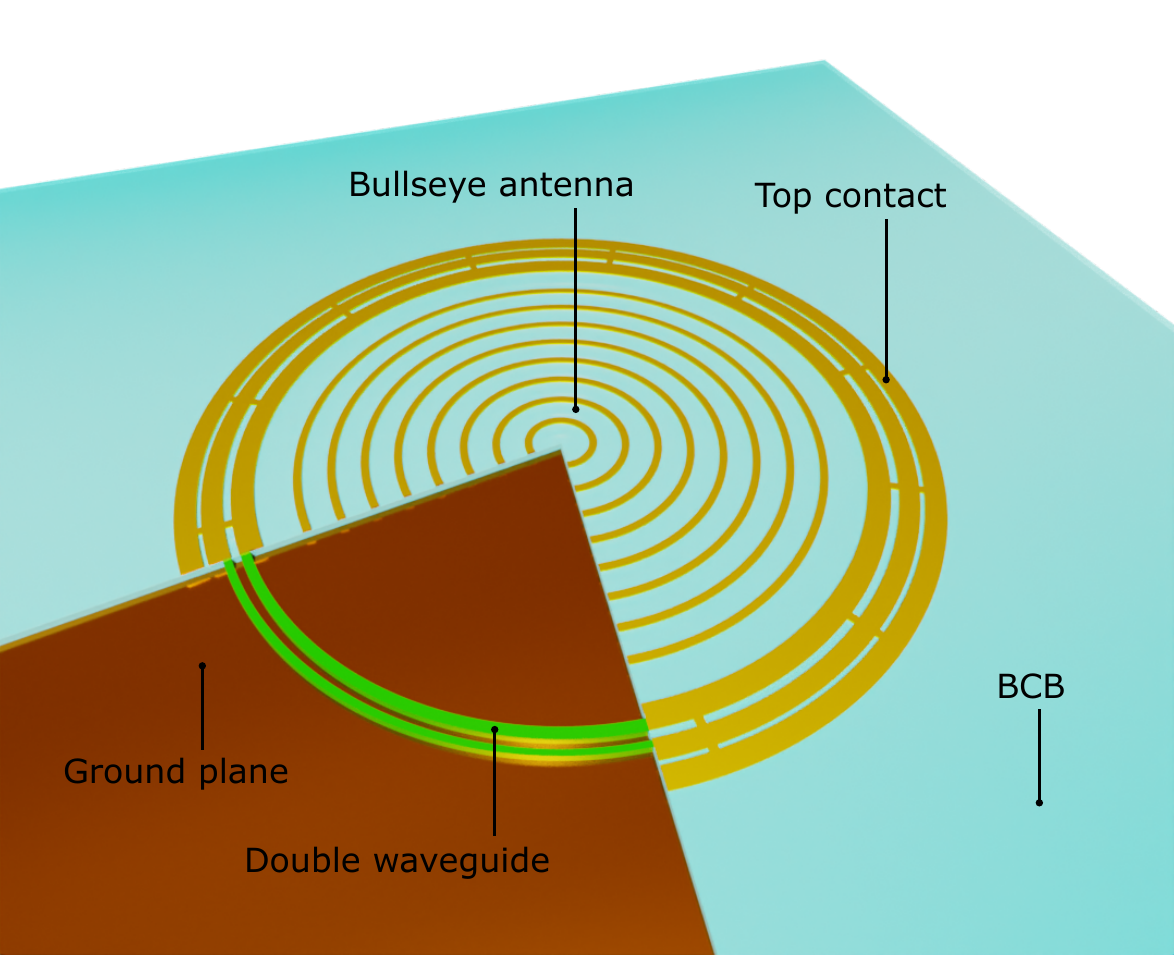}
\hspace{0.1\linewidth}
\includegraphics[width=0.3\linewidth]{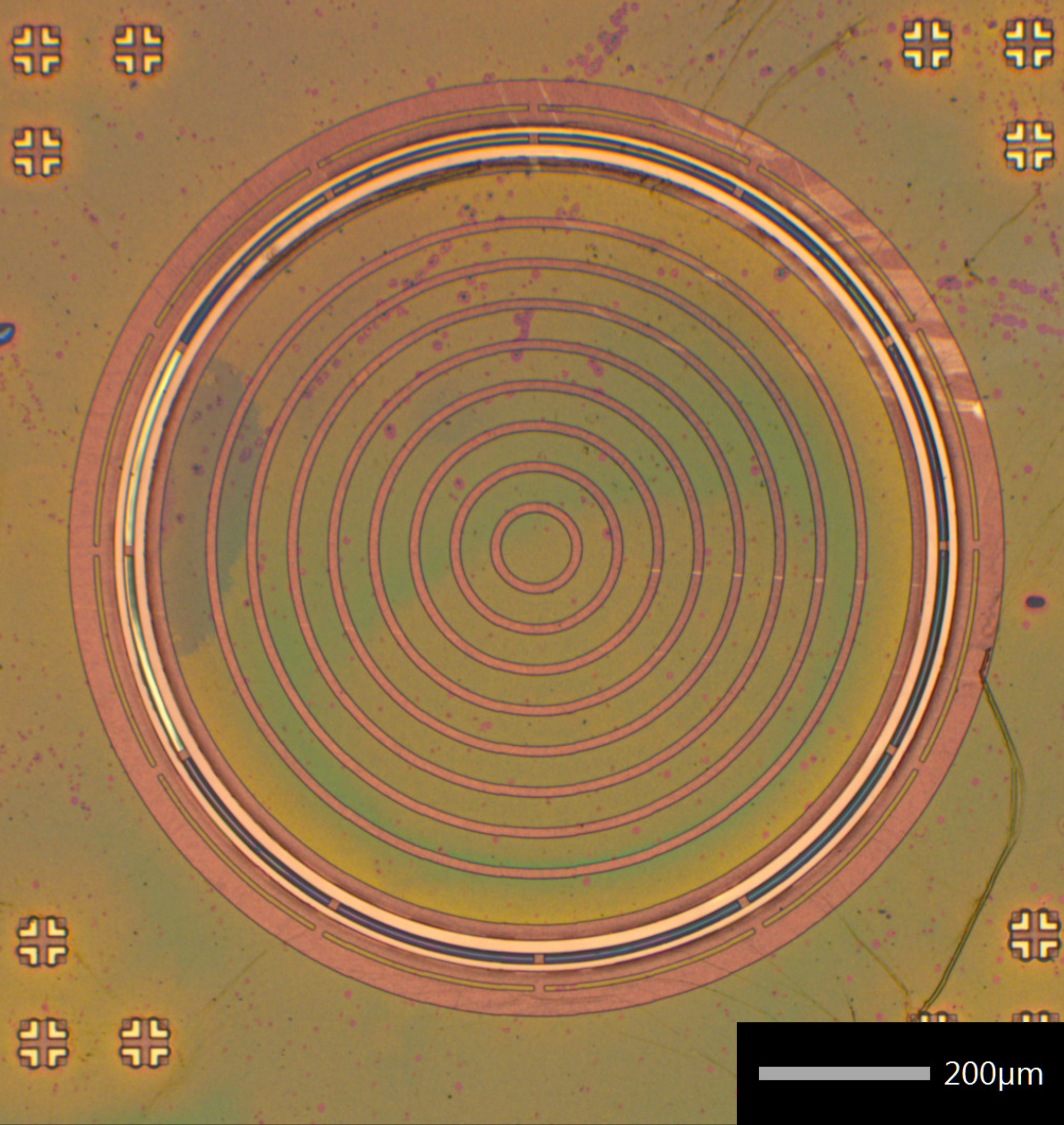}
\caption{(Left) Illustration of the device layout. (Right) Optical microscope image of a fully processed device.}
\label{sfig:schematic}
\end{figure}

The data presented in this paper were measured on two identical devices from the same
active region and processing batch. Their design is shown in Fig.\ref{sfig:schematic}, together with a microscope image of a fully processed device, and consists of a couple of concentric circular active waveguides, separated by a \SI{20}{\micro\meter} wide gap, with different widths that vary along the cavity length.
The metallic structure deposited on the planarizing dielectric at the center of the device act as a bullseye antenna, out-coupling the light traveling in the rings. A similar device design allowed to observe the generation of solitons on this platform \cite{micheletti2023terahertz}. In fact, the double waveguide configurations creates a superposition of the individual guided modes, with the generation of two supermodes, the symmetric and antisymmetric, featuring normal and anomalous dispersion, respectively. The mode with highest overlap with the active region will experience lower losses and prevail on the other one. The antisymmetric mode will naturally feature a higher overlap factor having a node in between the two ring waveguides where no active material is present. If the anomalous dispersion induced by the geometry is large enough to compensate for the other sources of chromatic dispersion, the propagation of a stable optical pulse in the cavity can be obtained, in which case soliton generation is observed.
In this work, we reduce the contribution of the anomalous dispersion, by tapering the waveguides widths, to simply reduce the overall dispersion and favor the generation of a stable quantum walk (QW) comb.

\subsection{Experimental Methods}
For all of the measurements presented, the device was mounted on a cold finger cryostat and cooled to an operating temperature of 40 K using a liquid helium transfer line and stabilizing the heat sink temperature with a temperature controller. \\
The laser was operated in continuous-wave mode, using a Keithley 2420 source. Additional microwave modulation was provided using a Rohde\&Schwarz SMB 100A microwave signal generator in combination with a Mini-Circuits ZVE-3W-183+ microwave amplifier (5.8-18 GHz) or a Pasternack PE15A5017 RF amplifier (0.7-6 GHz). For two-tone modulation, an additional Rohde\&Schwarz SGS100A SGMA RF source was used, and the two microwave signals combined together before the amplification stage.\\
The laser emission spectra were obtained using a Bruker Vertex 80v, a Fourier transform infrared spectrometer (FTIR) with a room-temperature deuterated triglycine sulfate (DTGS) detector. The optics and sample compartments of the FTIR were evacuated to avoid any THz absorption.\\
For SWIFT spectroscopy measurements \cite{burghoff2015evaluating}, a Schottky diode \cite{bozhkov2003semiconductor} was used as a fast detector at room temperature. The beatnote signal was collected at the FTIR output and measured with a Rohde\&Schwarz FSW67 RF spectrum analyzer with an IQ demodulator function. A detailed description of the experimental setup can be found in Ref.\cite{senica2022planarized}.

\newpage
\section{Supplementary text}
In the following, we present additional  results on the same devices presented in the main manuscript.

\subsection{Free-running operation}
The devices investigated in this work were first characterized under free-running operating conditions with a continuous-wave source. The optical spectra measured across the full dynamic range of the laser show a clear single-mode operation through the whole range. The two ring QCLs both show a similar gain competition mechanism for which the single-mode frequency jumps suddenly at certain values of the biase voltage.

\begin{figure}[h!]
\centering
\includegraphics[width=1\linewidth]{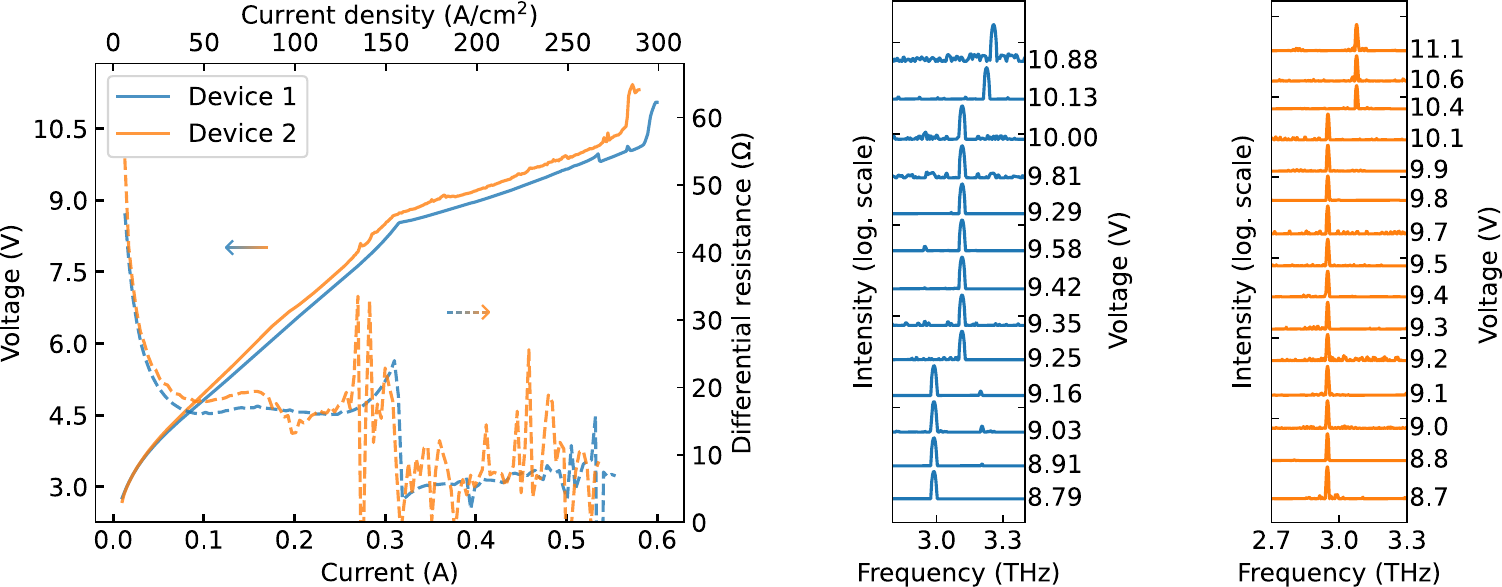}
\caption{Free-running characterization of the devices at 25 K under CW operating condition. Current-voltage (I-V) measurement of the two ring QCLs (a, solid line). The drop in the differential resistance (dashed line) indicates the lasing threshold, a clear signature of photon-assisted transport. Optical spectra of device 1 (b) and device 2 (c) measured under the same conditions. Both the devices jump to different lasing modes across the dynamic range, attributed to gain competition.}
\label{sfig:characterization}
\end{figure}

\subsection{Quantum walk comb regime}
We perform SWIFT measurements to observe the typical features of the QW comb formation regime. We apply an external RF modulation to the QCL device and tune it in resonance with the cavity roundtrip frequency. We observe a significant spectral broadening, and the reconstructed time evolution of the instantaneous frequency matches the predicted theoretical behavior. It changes progressively from a sinusoidal modulation, in the off-resonance case, to a clear half-cosine modulation, in the on-resonance one.

\begin{figure}[h!]
\centering
\includegraphics[width=0.9\linewidth]{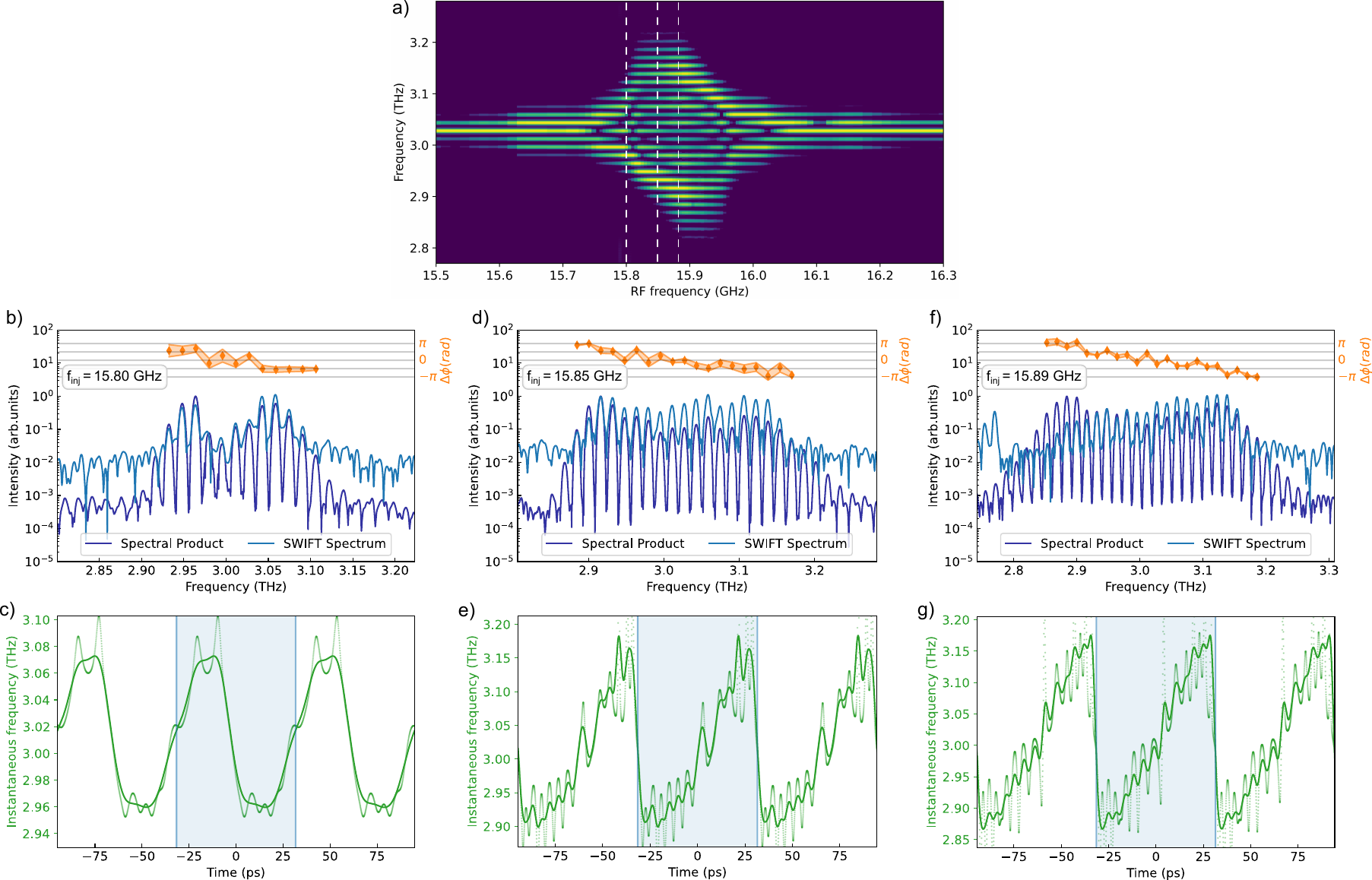}
\caption{Quantum walk comb regime. The expansion map (a) shows the spectral broadening when modulating the laser across the cavity resonance. The SWIFT measurements (b-g) show the progressive evolution of the instantaneous frequency profile in time, from a off-resonant nearly-sinusoidal trace (c) to a on-resonant half-cosine one (g) over the cavity round trip time.}
\label{sfig:QWexpansion}
\end{figure}

\newpage
\subsection{Subharmonic RF injection maps}

\begin{figure}[h!]
\centering
\includegraphics[width=1\linewidth]{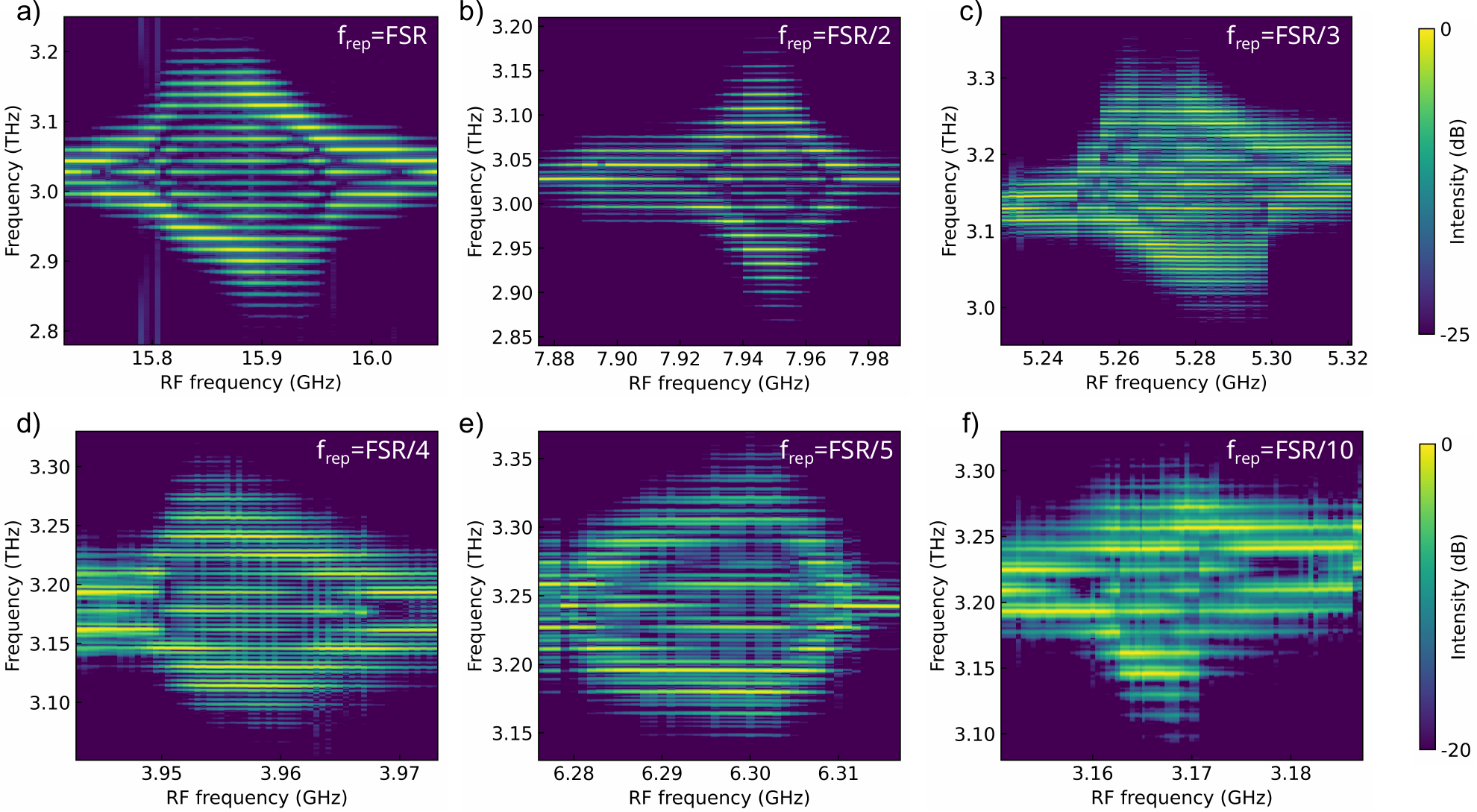}
\caption{Spectrum expansion maps as a function of the injected microwave frequency centered around subharmonics of the cavity FSR, indicated in each map. The mode spacing can be controlled and set in a range from 16 GHz (a) down to 1.6 GHz (f). A period-doubling dynamic appears in the measurements (e-f), for which the laser locks at half the injected RF frequency. }
\label{sfig:subharmonicMaps}
\end{figure}

The characteristic spectral broadening of the QW comb is also observed around subharmonics of the cavity FSR. In Fig.\ref{sfig:subharmonicMaps}, we report the measured optical spectra when driving the device with modulation frequencies ranging from the cavity FSR, down to its fourth subharmonic. The mode spacing can be tuned in a larger range as a consequence of the device electrical nonlinearity, which leads to a period-doubling dynamics of the microwave propagating in the ring. This effect is treated in more detail in the next section. 

\newpage
\subsection{Electrical nonlinearity leading to period-doubling}

\begin{figure}[b!]
\centering
\includegraphics[width=1\linewidth]{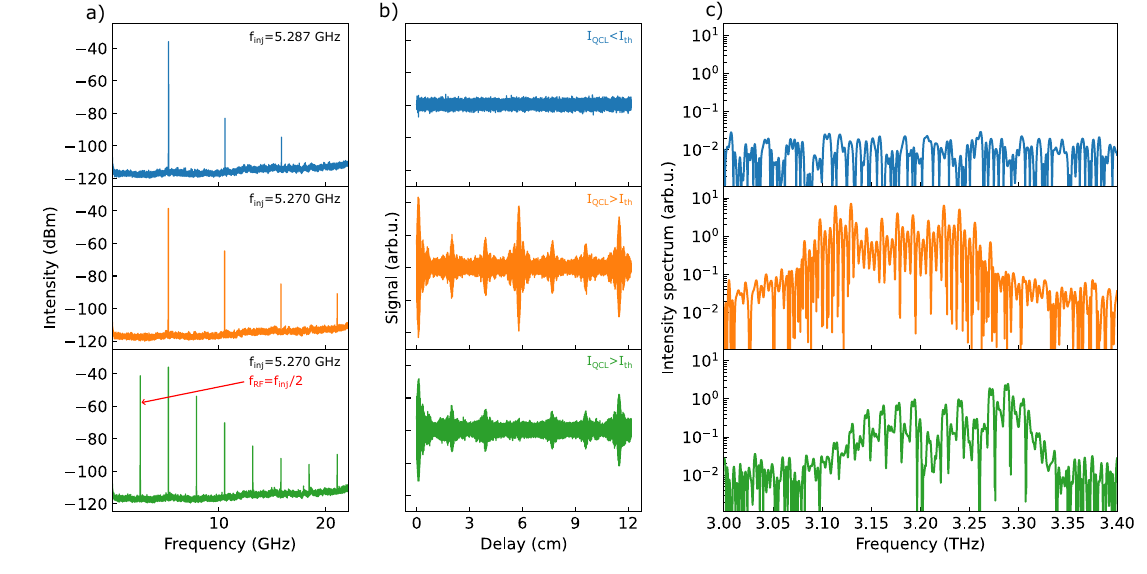}
\caption{Frequency division of the external microwave modulation. Electrical RF signals (a), interferograms (b), and optical spectra (c) measured with the QCL device operated at 40 K under microwave injection below (blue) and above (orange, green) the lasing threshold. The RF injection frequency is chosen to be close to the $1/3$ subharmonic of the fundamental cavity FSR. By increasing the RF power a strong electrical signal at half the injected frequency appears, modifying the spectrum FSR to $1/6$ of the cavity roundtrip frequency.}
\label{sfig:FreqDiv}
\end{figure}

\begin{figure}[t!]
\centering
\includegraphics[width=0.5\linewidth]{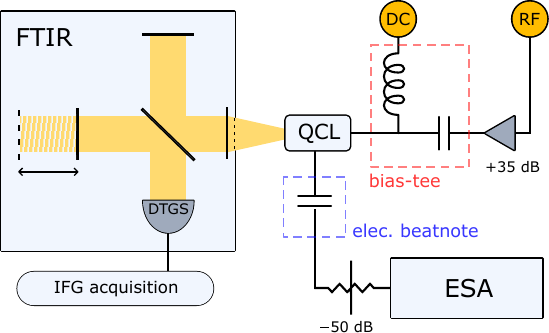}
\caption{Frequency division measurement set-up. The device is biased in DC with additional RF modulation. From a second QCL contact the electrical microwave signals is attenuated (-50 dB) and measured with an electrical spectrum analyser (ESA). The optical spectrum of the QCL is measured with a commercial FTIR.}
\label{sfig:FreqDivsetup}
\end{figure}

We measure the electrical response of the device under external microwave modulation at one third of the cavity FSR (Fig.\ref{sfig:FreqDiv}). With the QCL operated below its lasing threshold we already observe some harmonics of the injected RF tone being generated in the device. Above the lasing threshold the microwave modulation is imprinted on the optical spectrum. The nonlinear generation of the resonant component drives the spectrum expansion from a single-mode to a broadband state, while the injected frequency forces the mode spacing. The proliferation of optical modes reinforces the readout of the electrical signals because of the beating of the lasing modes in the active material. By setting an optimal combination of driving parameters (DC bias, RF frequency and power), a period-doubling dynamics can be observed, with the additional generation of the subharmonic of the injected RF signal and its odd-harmonics, populating the electrical spectrum measured from the QCL bias line (see Fig.\ref{sfig:FreqDivsetup}). This leads to a further modification of the optical spectrum, as highlighted by the measurements.


\section{Optical beat note detection}
\begin{figure}[h!]
\centering
\includegraphics[width=0.8\linewidth]{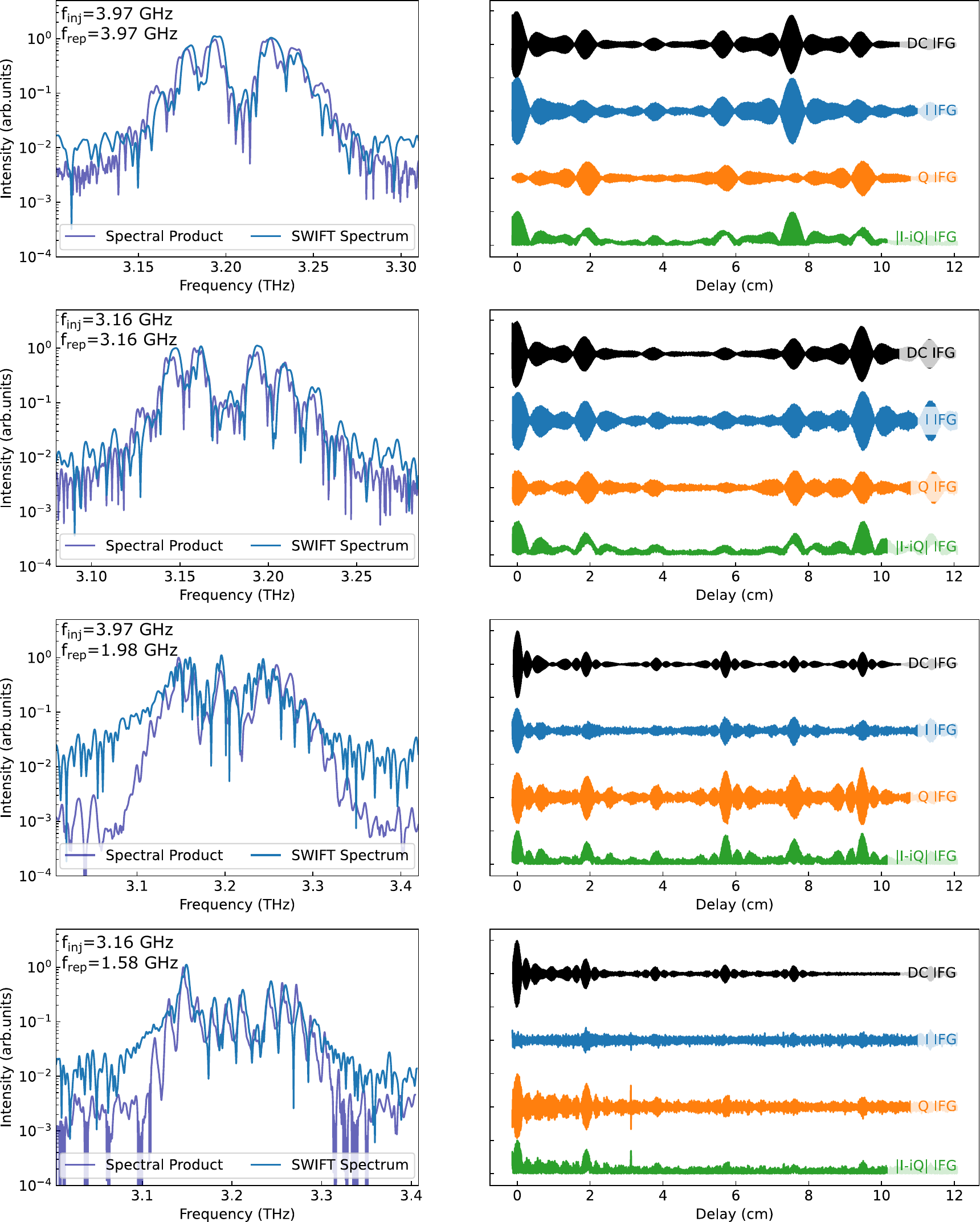}
\caption{Measurement of the laser spectrum with a fast Schottky detector. The detected signal is demodulated at the comb repetition rate corresponding to 1/4, 1/5, 1/8 and 1/10 of the cavity FSR from top to bottom respectively. The measurement resolution does not allow to resolve the individual modes and retrieve the intermodal phase difference needed for the SWIFT analysis.}
\label{sfig:optical_beat}
\end{figure}

In the main text we show the results of the SWIFTS measurements on the 1/2 and 1/3 subharmonic states of the THz ring QW comb laser. In this section, we demonstrate the coherence of the subharmonic states down to the 1/10 state. The resolution required to perform the SWIFT analysis goes beyond the capabilities of the FTIR used for this measurement, therefore, we use the fast detector in combination with the internal lock-in of the ESA to perform optical beat note interferometry of the laser output. The results are shown in Fig.\ref{sfig:optical_beat}. We observe a good agreement between the spectral product envelope and the SWIFT spectrum.

\newpage
\subsection{Theoretical Model}

In this subsection, we summarize the relevant theory used in modeling the quantum walk and subharmonic comb formation in the considered THz ring QCL. We begin with an overview of the master equation formalism, including the linewidth enhancement factor (LEF), which has been thought to play a critical role in enabling the phase modulation required for quantum walk comb formation. We then comment on recently observed pitfalls of this model and how they may be circumvented through consideration of an intensity-dependent LEF, which we argue is consistent with models leveraging the Bloch gain formalism. Finally, we introduce the nonlinear transmission line approach used in modeling the microwave propagation along the ring resonator. The discussion is informed by the results presented in~\cite{heckelmann2023quantum,Dikopoltsev2025theoryQW,opavcak2021frequency,opavcak2019theory,BurghoffOptica2020,opavcak2022origin}.

\subsubsection{Master equation for a ring QCL with a fast gain}

The master equation is derived from the Maxwell-Bloch formalism by considering the interaction of an electric field with a two-level system using the density matrix formalism. We use subscripts $u$ and $l$ to denote the upper and lower states, respectively. We write the Hamiltonian, $\hat{H}$, for the system in the interaction picture as $\hat{H} = \hat{H}_{0} + \hat{H}^{I}$ with $\hat{H}^{I} = -\hat{\mu}E$, where $\hat{\mu}$ is the dipole moment operator and $E$ the electric field. Then, the full Hamiltonian of the system can be written as

\begin{equation}\label{eq:Hamiltonian}
    \hat{H} = \begin{bmatrix}
        W_{l} & -\mu_{lu}E\\ -\mu_{lu}E & W_{u}
    \end{bmatrix} = \begin{bmatrix}
        \hbar\omega_{l} & -\mu_{lu}E \\ -\mu_{ul}E & \hbar\omega_{u}
    \end{bmatrix}.
\end{equation}

From this, the time evolution of the system is given by the von Neumann equation, which we write as

\begin{multline}
    \frac{d\hat{\rho}}{dt} = \frac{i}{\hbar}[\hat{H},\hat{\rho}] \\ = \frac{i}{\hbar}\begin{bmatrix}
        -\mu_{lu}\rho_{ul}E + \mu_{ul}\rho_{lu}E & \hbar\rho_{lu}(\omega_{l}-\omega_{u})-E\mu_{lu}(\rho_{uu}-\rho_{ll}) \\ \hbar\rho_{ul}(\omega_{u}-\omega_{l})-E\mu_{ul}(\rho_{ll}-\rho_{uu}) & -\mu_{ul}\rho_{lu}E + \mu_{lu}\rho_{ul}E
    \end{bmatrix}.
\end{multline}

To proceed, we take $\mu_{ul} = \mu_{lu} = \mu$, define $\omega_{0} = \omega_{u} - \omega_{l}$, and multiply by the total carrier number, $n_\mathrm{tot}$, such the number of carriers in a particular state $j$ are given by $n_{j} = \rho_{j}n_\mathrm{tot}$. We furthermore note that $\rho_{ul} = \rho_{lu}^{*}$ and simplify the subscripts $uu \rightarrow u$ and $ll \rightarrow l$. With these simplifications, the rate equations for the populations $n_{u}(z,t)$, $n_{l}(z,t)$, and $n_{ul}(z,t)$ become

\begin{subequations}
    \begin{equation}\label{eq:n_l1}
        \frac{\partial n_{l}}{dt} = 2\frac{\mu E}{\hbar}\mathrm{Im}
        \left\{{n_{ul}}\right\} + \frac{n_{u}}{T_{ul}} - \frac{n_{l}}{T_{lg}}+J_{t} +D\frac{\partial^{2}n_{l}}{\partial z},
    \end{equation}
    \begin{equation}\label{eq:n_u1}
        \frac{\partial n_{u}}{\partial t}=-2\frac{\mu E}{\hbar}\mathrm{Im}\left\{n_{ul}\right\}-\frac{n_{u}}{T_{ul}}-\frac{n_{u}}{T_{ug}} + J + D\frac{\partial^{2}n_{u}}{\partial z^2},
    \end{equation}
    \begin{equation}\label{eq:n_ul1}
        \frac{\partial n_{ul}}{dt} = in_{ul}\omega_{0} + i\frac{\mu E}{\hbar}(n_{u}-n_{l}) - \frac{n_{ul}}{T_{2}},
    \end{equation}
\end{subequations}

\noindent where we have introduced the carrier diffusion coefficient, $D$, lifetimes for dephasing processes, $T_{ij}$, where $i,j \in \{u,l,g\}$ (denoting upper, lower, and ground states), polarization dephasing time, $T_{2}$, pump current density, $J$, and thermal current density, $J_{t}$, which accounts for thermal excitation of carriers into the lower state.

The QW comb dynamics are generally considered to be a consequence of gain modulation coupling into phase modulation. This phenomenon is often described by the LEF, $\alpha(\omega)$, which characterizes the gain asymmetry or, likewise, coupling between changes in the gain and refractive index in the laser medium. Mathematically, the LEF is defined as

\begin{equation}
    \alpha = -\frac{\partial \chi_{R}/\partial N}{\partial \chi_{I}/\partial N}.
\end{equation}

Here, $\chi_{R}$ and $\chi_{I}$ represent the real and imaginary parts of the linear susceptibility, which are in turn related through the Kramers-Kronig relations. To understand how the LEF modifies our system of equations, we must find an expression for the susceptibility $\chi$ of the two-level system under consideration. We begin by noting that the macroscopic polarization $P$ is given by $P = \frac{n_{tot}}{L_{p}}\mathrm{Tr}{[\hat{\rho}\hat{\mu}]}$, where $L_{p}$ is the appropriate length such that $N_{tot} = n_{tot}/L_{p}$ gives the carrier density, and $\varepsilon_{0}$ is the permittivity of free space. From this, we find

\begin{equation}\label{eq:macro_pol}
    P(t) = \frac{\mu}{L_{p}}\bigl(n_{ul}(t)+n_{ul}^{*}(t)\bigr).
\end{equation}

The susceptibility is related to $P$ as $P = \varepsilon_{0}\chi E$. Taking $n_{u}$ and $n_{l}$ as constant, Eq. \ref{eq:n_ul1} may be solved in the Fourier domain and combined with Eq. \ref{eq:macro_pol} to yield

\begin{equation}\label{eq:susc_twolevel}
    \chi(\omega) = \frac{\mu^2}{L_{p}\varepsilon_{0}\hbar}\frac{n_{u}-n_{l}}{(\omega - \omega_{0}-i\gamma)} + \frac{\mu^2}{L_{p}\varepsilon_{0}\hbar}\frac{n_{u}-n_{l}}{(-\omega - \omega_{0}+i\gamma)},
\end{equation}

\noindent where $\gamma = 1/T_{2}$.

Following refs. \cite{opacak_theory_2019,bowden1993generalized,prati2007long,columbo2018dynamics}, inclusion of the LEF into our model can be achieved by considering only the LEF at the peak value of the gain, $\tilde{\alpha} = \alpha(\omega = \omega_{p})$, where $\omega_{p}$ is the frequency at which the peak gain occurs. Note that, with these definitions, $\omega_{p} = \omega_{0} + \frac{\tilde{\alpha}}{T_{2}}$. The modified linear susceptibility, $\chi$, between the upper and lower levels of the laser transition is then given as

\begin{equation}\label{eq:susc_LEF}
    \chi(\omega) = \frac{\mu^2(n_{u}-n_{l})}{L_{p}\varepsilon_{0}\hbar}\frac{(1+i\tilde{\alpha})^{2}}{(\omega-\omega_{0}-i\gamma)} + \frac{\mu^2(n_{u}-n_{l})}{L_{p}\varepsilon_{0}\hbar}\frac{(1-i\tilde{\alpha})^{2}}{(-\omega-\omega_{0}+i\gamma)}.
\end{equation}

By comparing Eq. \ref{eq:susc_LEF} with Eq. \ref{eq:susc_twolevel}, we find that inclusion of the LEF can be achieved in our system through modification of Eq. \ref{eq:n_ul1} to

\begin{equation}\label{eq:n_ul_LEF}
    \frac{\partial n_{ul}}{\partial t} = (i\omega_{0} - \frac{1}{T_{2}})n_{ul} + i\frac{E}{\hbar}\mu(1+i\tilde{\alpha})^{2}(n_{u}-n_{l}).
\end{equation}

Next, we must find a wave equation for the optical field. In a dispersive medium, where the permittivity is allowed to vary in frequency, the wave equation is given by

\begin{equation}
    \frac{\partial^2 E}{\partial z^2} - \frac{1}{c^2}\frac{\partial^2}{\partial t^2}\int_{-\infty}^{t}\epsilon_{r}(t-\tau)E(z,\tau)d\tau = \frac{1}{\varepsilon_{0}c^2}\frac{\partial^2P(z,t)}{\partial t^2}.\label{eq:WaveGenTD}
\end{equation}

This can be rewritten succinctly in the frequency domain as

\begin{equation}\label{eq:WaveGenFD}
    \frac{\partial^2\tilde{E}(z,\omega)}{\partial z^2}+k^2(\omega)\tilde{E}(z,\omega)=-\frac{\omega^2}{\varepsilon_{0}c}\tilde{P}(z,\omega),
\end{equation}

\noindent where tildes have been used to indicate frequency-domain quantities (omitted for constants and quantities such as $k(\omega) = \frac{\omega^2}{c^2}\tilde{\epsilon}_{r}(\omega)$ which will be expressed only as a function of frequency).

We further assume that in the ring resonator under consideration, the light will undergo unidirectional propagation, and we re-express $E$ and $n_{ul}$ as

\begin{subequations}
    \begin{equation}\label{eq:EtoA}
        E(z,t) = \frac{1}{2}(A(z,t)e^{i(\omega_{p}t - k_{p}z)}+A^{*}(z,t)e^{-i(\omega_{p}t-k_{p}z)}),
    \end{equation}
    \begin{equation}\label{eq:nultosigma}
        n_{ul}(z,t) = \sigma (z,t)e^{i(\omega_{p}t-k_{p}z)},
    \end{equation}
\end{subequations}

\noindent where we have taken the carrier to be at the peak of the gain, indicated by the subscript $p$. In frequency domain, then, we have

\begin{subequations}
    \begin{equation}
        \tilde{E}(z,\omega) = \frac{1}{2}(\tilde{A}(z,\omega-\omega_{p})e^{-ik_{p}z}+\tilde{A}^{*}(z,\omega+\omega_{p})e^{ik_{p}z}).
    \end{equation}
    \begin{equation}
        \tilde{n}_{ul}(z,\omega) = \tilde{\sigma}(z,\omega-\omega_{p})e^{-ik_{p}z}.
    \end{equation}
\end{subequations}

Plugging into Eq. \ref{eq:WaveGenFD} yields

\begin{multline}\label{eq:WaveEq_Expanded}
    \frac{1}{2}e^{-ik_{p}z}\left[\frac{\partial^{2}}{\partial z^{2}}\tilde{A}(z,\Omega) - 2ik_{p}\frac{\partial}{\partial z}\tilde{A}(z,\Omega)-k_{p}^{2}\tilde{A}(z,\Omega)\right] \\ + \frac{1}{2}k(\omega)^{2}e^{-ik_{p}z}\tilde{A}(z,\Omega) + \text{c.c.} \\ = -\frac{\omega^{2}\Gamma\mu}{L_{p}\varepsilon_{0}c^{2}}\tilde{\sigma}(z,\Omega)e^{-ik_{p}z} + \text{c.c.},
\end{multline}

\noindent where we have inserted the confinement factor, $\Gamma$, which accounts for the overlap between the electric field and the laser material, and introduced $\Omega = \omega - \omega_{p}$. Next, we Taylor expand $k(\omega)$ about $\omega_{p}$, which yields

\begin{equation}
    k(\omega)^{2} = k_{p}^{2} + 2k_{p}k'(\omega_{p})\Omega + k'(\omega_{p})^{2}\Omega^{2} + k_{p}k''(\omega_{p})\Omega^{2} + \mathcal{O}(\Omega^{3}).
\end{equation}

Here, $k'(\omega)$ represents $\frac{dk}{d\omega}$ and so on. Multiplying Eq. \ref{eq:WaveEq_Expanded} through by $e^{ik_{p}z}$, ignoring the fast-oscillating terms under the rotating wave approximation, plugging in the truncated series for $k(\omega)^{2}$ up to second order in $\Omega$, and taking the Fourier transform with respect to $\Omega$, we arrive at

\begin{multline}
    \frac{i}{2k_{p}}\biggl(\frac{\partial^{2}}{\partial z^{2}} - (k_{p}^{(1)})^{2}\frac{\partial^{2}}{\partial t^{2}}\biggr)A + \biggl(\frac{\partial}{\partial z} + k_{p}^{(1)}\frac{\partial}{\partial t}\biggr)A - i\frac{1}{2}k_{p}^{(2)}\frac{\partial ^2A}{\partial t^{2}} \\ = -i\frac{\Gamma \mu}{L_{p}\varepsilon_{0}c^{2}k_{p}}\biggl(\omega_{p}^{2}-2i\omega_{p}\frac{\partial}{\partial t}-\frac{\partial^{2}}{\partial t^{2}}\biggr)\sigma(z,t).
\end{multline}

Here, $k_{p}^{(1)} = k'(\omega_p)$ and, likewise, $k_{p}^{(2)} = k''(\omega_{p})$ Applying the slowly varying envelope approximation and neglecting the temporal derivatives of $\sigma$, which are relatively small contributions compared to the leading term, we arrive at the following system of equations describing the ring QCL dynamics,

\begin{subequations}
    \begin{equation}\label{eq:WaveEq_SVEA}
        \biggl(\frac{\partial}{\partial z} + k_{p}^{(1)} \frac{\partial}{\partial t}\biggr)A - i\frac{k_{p}^{(2)}}{2}\frac{\partial^{2} A}{\partial t^{2}} = -i\frac{\Gamma\mu\omega_{p}}{L_{p}\varepsilon_{0}cn_{p}}\sigma,
    \end{equation}
    \begin{equation}\label{eq:sigma_simp}
        \frac{\partial \sigma}{\partial t} = -\frac{1+i\tilde{\alpha}}{T_{2}}\sigma+i\mu(n_{u}-n_{l})(1+i\tilde{\alpha})^{2}\frac{A}{2\hbar},
    \end{equation}
    \begin{equation}\label{eq:n_u_simp}
        \frac{\partial n_{u}}{\partial t} = J - \frac{1}{T_{1}}n_{u} + i\frac{\mu}{2\hbar}(A^{*}\sigma - A\sigma^{*}),
    \end{equation}
    \begin{equation}\label{eq:n_l_simp}
        \frac{\partial n_{l}}{\partial t} = J_{t} + (\frac{1}{T_{1}}-\frac{1}{T_{ug}})n_{u}-\frac{1}{T_{lg}}n_{l}-i\frac{\mu}{2\hbar}(A^{*}\sigma - A\sigma^{*}).
    \end{equation}
\end{subequations}

From Eqns. \ref{eq:n_ul_LEF} and \ref{eq:n_l1}-\ref{eq:n_u1} to Eqns. \ref{eq:sigma_simp}-\ref{eq:n_l_simp}, we have assumed that diffusion is negligible and further made appropriate substitutes for $A$ and $\sigma$ using Eqns. \ref{eq:EtoA}-\ref{eq:nultosigma}.

Arrival at the master equation requires a few further approximations. Firstly, we assume efficient carrier extraction from the lower level such that the lower state population can be ignored ($n_{u} \approx 0$). Next, we assume a fast gain. Specifically, we assume that $T_{1}$ is much faster than any other timescale in the $n_{u}$ equation, such that $n_{u}$ may be adiabatically eliminated, which involves approximating it as its steady-state value. Specifically, through setting Eqns. \ref{eq:sigma_simp} and \ref{eq:n_u_simp} to zero, we find

\begin{equation}\label{eq:nu_adiab}
    n_{u} = \frac{T_{1}J}{1+|A|^{2}/A_\text{sat}^{2}},
\end{equation}

\noindent where $A_\text{sat}^{2} = \frac{2\hbar^{2}}{\mu^{2}T_{1}T_{2}}$ defines the saturation value for the electric field amplitude.

Next, we wish to solve for $\sigma$. This is most easily achieved in the frequency domain. Taking the Fourier transform with respect to $\Omega$ and using that $\omega_{p} = \omega_{0} + \frac{\tilde{\alpha}}{T_{2}}$ yields

\begin{equation}
    \tilde{\sigma} = i\frac{\mu}{2\hbar}T_{2}(1+i\tilde{\alpha})^{2}\frac{1}{1+iT_{2}(\omega-\omega_{0})}\mathcal{F}_{\Omega}\{An_{u}\}.  
\end{equation}

Taylor expanding the fraction to second order gives

\begin{equation}
    \tilde{\sigma} \approx i\frac{\mu}{2\hbar}T_{2}(1+i\tilde{\alpha})\biggl(1 - i\frac{T_{2}}{1+i\tilde{\alpha}}(\omega-\omega_{p}) - \frac{T_{2}^{2}}{(1+i\tilde{\alpha})^{2}}(\omega-\omega_{p})^{2}\biggr)\mathcal{F}_{\Omega}\{An_{u}\}.
\end{equation}

Finally, letting $\tilde{T}_{2} = \frac{T_{2}}{1+i\tilde{\alpha}}$ and neglecting terms containing derivatives of $n_{u}$, which is consistent with the above adiabatic approximation for $n_{u}$, we find

\begin{equation}\label{eq:sigma_expression}
    \sigma \approx i(1+i\tilde{\alpha})\frac{\mu T_{1}T_{2}}{2\hbar}\frac{J}{1+|A|^{2}/A_\text{sat}^{2}}\biggl(A-\tilde{T}_{2}\frac{\partial A}{\partial t} + \tilde{T}_{2}\frac{\partial^{2}A}{\partial t^{2}}\biggr),
\end{equation}

\noindent where we have also plugged in the expression for $n_{u}$ from Eq. \ref{eq:nu_adiab}.

At last, by combining Eq. \ref{eq:sigma_expression} with \ref{eq:WaveEq_SVEA} we arrive at the master equation for the evolution of the field of a fast gain QCL, including the LEF,

\begin{multline}\label{eq:mastereq_LEF}
    \biggl(\frac{\partial}{\partial z} + k_{p}^{(1)}\frac{\partial}{\partial t}\biggr)A = i\frac{k_{p}^{(2)}}{2}\frac{\partial^{2}A}{\partial t^{2}} \\ + (1 + i\tilde{\alpha})\frac{\Gamma\mu^{2}\omega_{p}T_{1}T_{2}}{2n\varepsilon_{0}cL_{p}\hbar}\frac{J}{1+|A|^{2}/A_\text{sat}^{2}}\biggl(A-\tilde{T}_{2}\frac{\partial A}{\partial t} + \tilde{T}_{2}^{2}\frac{\partial^{2}A}{\partial t^{2}}\biggr) - \frac{\alpha_{c}}{2}A,
\end{multline}

\noindent where we have additionally inserted the loss term, $\alpha_{c}$, which accounts both for propagation loss and outcoupling losses. It is typical to then define the gain $g(I) = \frac{\Gamma\mu^{2}\omega_{p}T_{1}T_{2}}{n\varepsilon_{0}cL_{p}\hbar}\frac{J}{1+|A|^{2}/A_\text{sat}^{2}} = \frac{g_{0}}{1+|A|^{2}/A_\text{sat}^{2}}$. However, under the considered case of microwave injection, we have that $J(z,t) = J_\text{DC} + J_\text{mod}(z,t)$, where we assume that the DC component of the current density is also spatially invariant. We thus define instead $g_{0} = \frac{\Gamma\mu^{2}\omega_{p}T_{1}T_{2}}{n\varepsilon_{0}cL_{p}\hbar}J_\text{DC}$ such that $g(I) = \frac{g_{0}(1+m(z,t))}{1+|A|^{2}/A_\text{sat}^{2}}$, with $m(z,t) = J_\text{mod}(z,t)/J_\text{DC}$. We consider the co-moving spatial coordinate $\zeta = \frac{1}{k_{p}^{(1)}}t + z$, which together allows us to rewrite Eq. \ref{eq:mastereq_LEF} as

\begin{equation}\label{eq:mastereq_LEF_comov}
    \frac{\partial A}{\partial \zeta} = i\frac{k_{p}^{(2)}}{2}\frac{\partial^{2}A}{\partial t^{2}} + (1 + i\tilde{\alpha})\frac{g_{0}(1 + m(\zeta-\frac{1}{k_{p}^{(1)}}t,t))}{2(1+|A|^{2}/A_\text{sat}^{2})}\biggl(A-\tilde{T}_{2}\frac{\partial A}{\partial t} + \tilde{T}_{2}^{2}\frac{\partial^{2}A}{\partial t^{2}}\biggr) - \frac{\alpha_{c}}{2}A.
\end{equation}

As has been previously noted in ref. \cite{letsou2026high}, however, Eq. \ref{eq:mastereq_LEF_comov} presents an issue in terms of the modeling of QW comb formation. Specifically, the dominant features of the QW comb arise due to synchronous phase modulation of the intracavity field in a fast gain laser. Equation \ref{eq:mastereq_LEF_comov} offers two primary mechanisms for the current modulation to couple into modulation of the phase of $A$. The first is for the current modulation to modify the amplitude of the field $A$ which then couples into a phase modulation through the effective Kerr nonlinearity arising from the interplay of gain saturation and the LEF. The second pathway is through the direct phase modulation arising from the product $ig_{0}\tilde{\alpha}mA$ on the right-hand side. At steady state, in the present form of Eq. \ref{eq:mastereq_LEF_comov}, these two phase contributions exactly cancel one another out. To show this, let us consider these two terms in isolation. We have that

\begin{equation}\label{eq:phasecontributions}
    \frac{\partial A}{\partial \zeta} \propto i\tilde{\alpha}\frac{g_{0}}{2}(1+m)A - \frac{i\tilde{\alpha}g_{0}(1+m)|A|^{2}}{2(A_\text{sat}^{2}+|A|^{2})}A,
\end{equation}

\noindent where we have used that $\frac{1}{1+|A|^{2}/A_\text{sat}^2}=1-\frac{|A|^{2}}{A_\text{sat}^2+|A|^{2}}$. The first term on the right-hand side is the previously described direct modulation term, while the second represents the effective Kerr nonlinearity. Let us compare the strength of these two effects by letting $m = 0$ and instead considering the impact of the modulation as a small perturbation $\Delta g_{0}$ to the steady-state solution under continuous drive $g_{0}$. The approximate steady-state signal amplitude, $A_{s}$, can be found from Eq. \ref{eq:mastereq_LEF_comov} as

\begin{equation}\label{eq:CW_SteadyState}
    |A_{s}|^{2} = A_\text{sat}^{2}\bigl(\frac{g_{0}}{\alpha_{c}}-1\bigr).
\end{equation}

Then, for a small perturbation to the gain, $\Delta g_0$, the corresponding change to the steady-state signal amplitude is

\begin{equation}
    \Delta |A_{s}^{2}| = A_\text{sat}^{2}\frac{\Delta g_{0}}{\alpha_\text{c}}.
\end{equation}

Let us now consider the change in the contributions from the two terms in Eq. \ref{eq:phasecontributions} under this perturbation. We write the change to $\frac{\partial A}{\partial \zeta}$, which we call $\Delta \frac{\partial A}{\partial \zeta}$, as

\begin{multline}
    \Delta \frac{\partial A}{\partial \zeta} \propto i\frac{\Delta g_{0}}{2}\tilde{\alpha}A_{s} - \frac{i\tilde{\alpha}g_{0}\Delta|A_{s}|^{2}}{2(A_\text{sat}^{2}+|A_{s}|^{2})}A_{s} \\ = i\frac{\Delta g_{0}}{2}\tilde{\alpha}A_{s} - \frac{i\tilde{\alpha}g_{0}A_\text{sat}^{2}\Delta g_{0}}{2\alpha_{c}\bigl(A_\text{sat}^{2}+A_\text{sat}^{2}(\frac{g_{0}}{\alpha_{c}}-1)\bigr)}A_{s} \\ = i\tilde{\alpha}\frac{\Delta g_{0}}{2}A_{s}-i\tilde{\alpha}\frac{\Delta g_{0}}{2}A_{s} = 0.
\end{multline}

Thus, at steady-state, these two dominant contributions to the phase cancel one another out as they are equal and opposite. In ref. \cite{letsou2026high}, it was suggested that this issue can be reconciled through inclusion of a large additional Kerr nonlinear term in the master equation. Here, we suggest that such an additional Kerr term may arise naturally by considering the origin of the LEF as arising from Bloch gain.

\subsubsection{Introduction of the Bloch gain}

To motivate the subsequent argument, let us consider in more detail how the discussed balancing of terms occurs. The issue arises since the real gain term, $\frac{g_{0}}{1+|A|^{2}/A_\text{sat}^{2}}$, shares the same saturation term (intensity dependence) as the imaginary contribution through the linewidth enhancement factor, $\frac{g_{0}\tilde{\alpha}}{1+|A|^{2}/A_\text{sat}^2}$. If, instead, the linewidth enhancement factor exhibited a distinct intensity dependence, the situation would be resolved. As an example, let us consider a linear dependence of $\tilde{\alpha}$ on the intensity such that we may express $\tilde{\alpha}(I) = \tilde{\alpha}_{0}(1+|A|^{2}/\tilde{A}_\text{sat}^2)$, where $\tilde{\alpha}_{0}$ is therefore the LEF at the peak value of the gain at $A = 0$ and $\tilde{A}_\text{sat}$ gives the linear coefficient, expressed as a saturation value of the field which is in general distinct from $A_\text{sat}$. Then, Eq. \ref{eq:mastereq_LEF_comov} becomes

\begin{multline}\label{eq:MEQ_satLEF_comov}
    \frac{\partial A}{\partial \zeta} = i\frac{k_{p}^{(2)}}{2}\frac{\partial^{2}A}{\partial t^{2}} \\ + \biggr(1 + i\tilde{\alpha}_{0}\bigl(1+\frac{|A|^{2}}{\tilde{A}_\text{sat}^{2}}\bigr)\biggl)\frac{g_{0}(1 + m(\zeta-\frac{1}{k_{p}^{(1)}}t,t))}{2(1+|A|^{2}/A_\text{sat}^{2})}\biggl(A-\tilde{T}_{2}\frac{\partial A}{\partial t} + \tilde{T}_{2}^{2}\frac{\partial^{2}A}{\partial t^{2}}\biggr) - \frac{\alpha_{c}}{2}A.
\end{multline}

Analyzing the dominant terms which couple current modulation into phase as before, we find

\begin{equation}\label{eq:phaseconts_satLEF}
    \frac{\partial A}{\partial \zeta} \propto i\frac{g_{0}}{2}\tilde{\alpha}_{0}(1+m)A - \frac{i\tilde{\alpha}_{0}g_{0}(1+m)|A|^{2}/A_\text{sat}^{2}}{2(1+|A|^{2}/A_\text{sat}^2)}A+\frac{i\tilde{\alpha}_{0}g_{0}(1+m)|A|^{2}/\tilde{A}_\text{sat}^{2}}{2(1+|A|^{2}/A_\text{sat}^2)}A.
\end{equation}

Compared to Eq. \ref{eq:phasecontributions}, we see that the linear dependence of $\tilde{\alpha}$ on $|A|^{2}$ has given rise to an additional saturating Kerr-like term which has an opposite sign compared to the original Kerr-like term arising from gain saturation. With $\tilde{A}_\text{sat}^{2} = A_\text{sat}^{2}$, the two terms exactly cancel each other out, leaving only the direct modulation term. However, more generally, the addition of this intensity dependence to $\tilde{\alpha}$ creates an imbalance between the phase contributions of the direct modulation and Kerr-like terms, which opens a pathway for the current modulation to couple into a phase modulation. To justify that we might expect such an intensity dependence of the LEF, we now turn to the Bloch gain formalism presented in ref. \cite{opavcak2021frequency}.

As previously described, the LEF has been introduced as a phenomenological factor to characterize the gain asymmetry of a laser. Bloch gain provides an explanation for the physical origin of this asymmetry. Bloch gain was predicted as a consequence of Bloch oscillations in a semiconductor superlattices~\cite{ktitorov1972bragg,sekine2005dispersive} and later generalized to semiconductor heterostructures more broadly~\cite{willenberg2003intersubband,wacker2002gain}, arising as a consequence of scattering-assisted optical transitions between subbands. In this framework, the gain profile of a semiconductor laser is thought to be comprised of both a harmonic oscillator component, which yields the usual symmetric Lorentzian gain lineshape, and a Bloch oscillator component, which yields the asymmetric Bloch gain contribution.

Formally, following ref. \cite{opavcak2021frequency}, we may write the gain $g(\omega)$ as

\begin{equation}\label{eq:Bloch_g}
    g(\omega) = g^{H}(\omega) + g^{B}(\omega),
\end{equation}

\noindent and, likewise, the susceptibility $\chi(\omega)$ may be rewritten as

\begin{equation}\label{eq:Bloch_chi}
    \chi(\omega) = \chi^{H}(\omega) + \chi^{B}(\omega).
\end{equation}

Here, superscripts $H$ and $B$ refer respectively to the harmonic oscillator and Bloch oscillator contributions. A simplified model for the susceptibilities $\chi^{H}$ and $\chi^{B}$ is presented in ref. \cite{opavcak2021frequency} by assuming that the electron distributions follow Boltzmann statistics. Additional approximations include assuming that the electron masses and scattering potentials of the upper and lower states are the same, and, correspondingly, subband nonparabolicity is neglected. Here, we quote the result,

\begin{subequations}
    \begin{equation}\label{eq:susc_H}
        \chi^{H}(\omega) = \frac{\mu^{2}\omega_{0}^{2}}{L_{p}\varepsilon_{0}\hbar\omega^{2}}\frac{(n_{u}-n_{l})}{(\omega-\omega_{0}-i\gamma)} + \frac{\mu\omega_{0}}{L_{p}\varepsilon_{0}\hbar}\frac{(n_{u}-n_{l})}{(-\omega-\omega_{0}+i\gamma)},
    \end{equation}
    \begin{equation}\label{eq:susc_B}
        \chi^{B}(\omega) = -i\frac{\mu^{2}\omega_{0}^{2}}{2L_{p}\varepsilon_{0}\hbar\omega^{2}}\frac{\gamma\bigl(1-e^{-\frac{\hbar|\omega-\omega_{0}|}{k_{B}T}}\bigr)(n_{u}+n_{l})}{|\omega-\omega_{0}|(\omega-\omega_{0}-i\gamma)} + i\frac{\mu^{2}\omega_{0}^{2}}{2L_{p}\varepsilon_{0}\hbar\omega^{2}}\frac{\gamma\bigl(1-e^{-\frac{\hbar|-\omega-\omega_{0}|}{k_{B}T}}\bigr)(n_{u}+n_{l})}{|-\omega-\omega_{0}|(-\omega-\omega_{0}+i\gamma)}.
    \end{equation}
\end{subequations}

The first contribution from the harmonic oscillator term is approximately equivalent to the previously defined susceptibility in Eq. \ref{eq:susc_twolevel} besides the near unity factor $\omega_{0}^{2}/\omega^{2}$. Notably, the Bloch contribution also contains a similar form besides the imaginary unit, a dependence on the sum $n_{u} + n_{l}$ of the populations instead of their difference, and some additional prefactors which we lump together in a single term, $b$, given as

\begin{equation}
    b(\omega) = -\frac{\gamma/2}{|\omega-\omega_{0}|}\bigl(1-e^{-\frac{\hbar|\omega-\omega_{0}|}{k_{B}T}}\bigr)\frac{n_{u}+n_{l}}{n_{u}-n_{l}}.
\end{equation}

With this definition, we may combine the two contributions into a single equation for the susceptibility to find

\begin{equation}\label{eq:totsusc_Bloch}
    \chi(\omega) = \frac{\mu^{2}\omega_{0}^{2}}{L_{p}\varepsilon_{0}\hbar\omega^{2}}\frac{(n_{u}-n_{l})}{\omega-\omega_{0}-i\gamma}(1+ib(\omega)) + \frac{\mu^{2}\omega_{0}^{2}}{L_{p}\varepsilon_{0}\hbar\omega^{2}}\frac{(n_{u}-n_{l})}{-\omega-\omega_{0}+i\gamma}(1-ib(-\omega)).
\end{equation}

By further considering that $|\omega - \omega_{0}| << k_{B}T/\hbar$, the factor $b$ may be further simplified as

\begin{equation}\label{eq:b_param}
    b = -\frac{\hbar\gamma}{2k_{B}T}\frac{n_{u}+n_{l}}{n_{u}-n_{l}}.
\end{equation}

By comparing Eq. \ref{eq:totsusc_Bloch} and Eq. \ref{eq:susc_LEF}, we see how the $b$ factor plays a similar role to the LEF in quantifying the gain asymmetry. Importantly for the present analysis, however, the explicit dependence of $b$ on the populations $n_{u}$ and $n_{l}$ allows direct calculations of the intensity and current dependence of $b$. These additional dependencies, as explored qualitatively through the toy model of Eq. \ref{eq:MEQ_satLEF_comov}, break the balance of phase contributions arising from the direct modulation and Kerr-like terms in the Bloch gain formalism.

We proceed to solve for this dependence, again carefully following ref. \cite{opavcak2021frequency}. We begin by writing expressions for for the population sum and difference terms, $n_{s} = n_{u} + n_{l}$ and $\Delta n = n_{u} - n_{l}$. From Eqs. \ref{eq:n_u1} and \ref{eq:n_l1}, we have

\begin{subequations}
    \begin{equation}
        \frac{\partial \Delta n}{\partial t} = J - J_{t} + n_{s}\bigl(\frac{1}{T_{\Delta}}-\frac{1}{T_{ul}}\bigr)-\Delta n\bigl(\frac{1}{T_{s}}+\frac{1}{T_{ul}}\bigr)-4\frac{\mu E}{\hbar}\text{Im}\{n_{ul}\},
    \end{equation}
    \begin{equation}
        \frac{\partial n_{s}}{\partial t} = J + J_{t} - \frac{n_{s}}{T_{s}}+\frac{\Delta n}{T_{\Delta}},
    \end{equation}
\end{subequations}

\noindent where we have, as before, neglected diffusion and introduced the lifetimes $T_{s} = \bigl(\frac{1}{2T_{ug}} + \frac{1}{2T_{lg}}\bigr)^{-1}$ and $T_{\Delta} = \bigl(\frac{1}{2T_{lg}} - \frac{1}{2T_{ug}}\bigr)^{-1}$.

To proceed, we make again make a similar ansatz for $E$ and $\sigma$ as in Eqs. \ref{eq:EtoA} and \ref{eq:nultosigma}, though this time taking $\omega_{0}$ as the carrier. We therefore have

\begin{subequations}
    \begin{equation}\label{eq:EtoA_Bloch}
        E(z,t) = \frac{1}{2}(A(z,t)e^{i(\omega_{0}t - k_{0}z)}+A^{*}(z,t)e^{-i(\omega_{0}t-k_{0}z)}),
    \end{equation}
    \begin{equation}\label{eq:nultosigma_Bloch}
        n_{ul}(z,t) = \sigma (z,t)e^{i(\omega_{0}t-k_{0}z)}.
    \end{equation}
\end{subequations}

With these substitutions, we find

\begin{subequations}
    \begin{equation}\label{eq:DelN_demod}
        \frac{\partial \Delta n}{\partial t} = J - J_{t} + n_{s}\bigl(\frac{1}{T_{\Delta}}-\frac{1}{T_{ul}}\bigr)-\Delta n\bigl(\frac{1}{T_{s}} + \frac{1}{T_{ul}}\bigr)+2\frac{\mu}{\hbar}\text{Im}\{A\sigma^{*}\},
    \end{equation}
    \begin{equation}\label{eq:ns_demod}
        \frac{\partial n_{s}}{\partial t} = J + J_{t} - \frac{n_{s}}{T_{s}}+\frac{\Delta n}{T_{\Delta}},
    \end{equation}
    \begin{equation}\label{eq:sigma_demod}
        \frac{\partial \sigma}{\partial t} = \frac{i\mu T_{2}}{2\hbar}A\Delta n - \frac{\sigma}{T_{2}}.
    \end{equation}
\end{subequations}

Solving for the steady-state values of $\sigma$, $\Delta n$, and $n_{s}$ in the adiabatic approximation yields the following expressions for $\Delta n$ and $n_{s}$,

\begin{subequations}
    \begin{equation}
        \Delta n = \frac{(J-\tilde{J}_{t})T_{\Delta n}}{1+|A|^{2}/\tilde{A}^{2}_\text{sat}},
    \end{equation}
    \begin{equation}
        n_{s} = T_{s}(J+\Upsilon\tilde{J}_{t})+\frac{(J-\tilde{J}_{t})T_{n_{s}}}{1+|A|^{2}/\tilde{A}_\text{sat}^{2}},
    \end{equation}
\end{subequations}

\noindent where

\begin{subequations}
    \begin{equation}
        \Upsilon = \biggl(1-T_{s}\bigl(\frac{1}{T_{\Delta}}-\frac{1}{T_{ul}}\bigr)\biggr)\biggl(1+T_{s}\bigl(\frac{1}{T_{\Delta}}-\frac{1}{T_{ul}}\bigr)\biggr)^{-1},
    \end{equation}
    \begin{equation}
        \tilde{J}_{t}=\frac{J_{t}}{\Upsilon},
    \end{equation}
    \begin{equation}
        T_{\Delta n} = \biggl(1+T_{s}\bigl(\frac{1}{T_{\Delta}}-\frac{1}{T_{ul}}\bigr)\biggr)\biggl(\frac{1}{T_{s}}+\frac{1}{T_{ul}}-\frac{T_{s}}{T_{\Delta}}\bigl(\frac{1}{T_{\Delta}}-\frac{1}{T_{ul}}\bigr)
        \biggr)^{-1},
    \end{equation}
    \begin{equation}
        T_{n_{s}}=\frac{T_{s}}{T_{\Delta}}T_{\Delta n},
    \end{equation}
    \begin{equation}
        \tilde{A}_{s}^{2} = \frac{\hbar^{2}}{\mu^{2}T_{2}}\biggl(\frac{1}{T_{s}}+\frac{1}{T_{ul}}-\frac{T_{s}}{T_{\Delta}}\bigl(\frac{1}{T_{\Delta}}-\frac{1}{T_{ul}}\bigr)\biggr).
    \end{equation}
\end{subequations}

Having now solved for the population sum and difference terms, we can explicitly write the dependence of $b$ on the intracavity intensity. Plugging into Eq. \ref{eq:b_param} gives

\begin{equation}\label{eq:b_intdep}
    b = -\frac{\hbar \gamma}{2k_{B}T}\biggl(\frac{T_{n_{s}}}{T_{\Delta n}}+\frac{(J+\Upsilon\tilde{J}_{t})T_{s}}{(J-\tilde{J}_{t})T_{\Delta n}}\bigl(1+\frac{|A|^{2}}{\tilde{A}_{s}^{2}}\bigr)\biggr).
\end{equation}

From this, we observe that $b$ exhibits a dominantly linear dependence on the intracavity intensity. There is a small additional dependence of the pump current; however, this dependence can be largely neglected if the thermal current density, $J_{t}$, is small. Derivation of a propagation equation including the Bloch gain follows a similar procedure to that laid out in the previous section. From the susceptibility in Eq. \ref{eq:totsusc_Bloch}, taking $\omega_{0}/\omega \approx 1$, we find the modified expression for $n_{ul}$ in the presence of the Bloch gain,

\begin{equation}
    \frac{\partial n_{ul}}{\partial t} = (i\omega_0 - \frac{1}{T_{2}})n_{ul}+i\frac{\mu E}{\hbar}(1+ib)\Delta n.
\end{equation}

Re-expressing $E$ and $n_{ul}$ as in Eqs. \ref{eq:EtoA} and \ref{eq:nultosigma} where now $\omega_{p}  = \omega_{0} + \frac{\xi}{T_{2}}$, with $\xi = (\sqrt{b^{2}+1}-1)/b$, we find under the slowly varying envelope approximation

\begin{subequations}
    \begin{equation}
        \biggl(\frac{\partial}{\partial z}-k_{p}^{(1)}\frac{\partial}{\partial t}\biggr)A - i\frac{k_{p}^{(2)}}{2}\frac{\partial^{2}A}{\partial t^{2}} = -i\frac{\Gamma\mu\omega_{p}}{L_{p}\varepsilon_{0}cn_{p}}\sigma,
    \end{equation}
    \begin{equation}
        \frac{\partial \sigma}{\partial t} = -\biggl(\frac{1+i\xi}{T_{2}}\biggr)\sigma + i\frac{\mu}{2\hbar}(1+ib)An_{u},
    \end{equation}
    \begin{equation}
        \frac{\partial n_{u}}{\partial t} = J - \frac{1}{T_{1}}n_{u} + i\frac{\mu}{2\hbar}(A^{*}\sigma - A\sigma^{*}),
    \end{equation}
\end{subequations}

\noindent where we, as before, assume a negligible lower state population. Solving for $n_{u}$ and $\sigma$ under the same set of assumptions as before gives the following propagation equation for the field of a fast gain laser with a Bloch gain contribution,

\begin{multline}\label{eq:mastereq_Bloch}
    \biggl(\frac{\partial}{\partial z}+k_{p}^{(1)}\frac{\partial}{\partial t}\biggr)A = i\frac{k_{p}^{(2)}}{2}\frac{\partial^{2}}{\partial t^{2}} \\ + \frac{\Gamma\mu^{2}\omega_{p}T_{1}T_{2}}{2\hbar L_{p}\varepsilon_{0}cn_{p}}\frac{(1+ib)}{(1+i\xi)}\frac{J}{1+|A|^{2}/A_\text{sat}^{2}}\biggl(A - \frac{T_{2}}{(1+i\xi)}\frac{\partial A}{\partial t}-\frac{T_{2}^{2}}{(1+i\xi)}\frac{\partial^{2}A}{\partial  t^{2}}\biggr) - \frac{\alpha_{c}}{2}A.
\end{multline}

As before, we have introduced the loss term, $\alpha_{c}$, and the saturation amplitude, here $A_\text{sat}^{2} = \frac{2\hbar(1+\xi^{2})}{\mu^{2}T_{1}T_{2}(1+\xi b)}$. Importantly, Eq. \ref{eq:mastereq_Bloch} is qualitatively similar to Eq. \ref{eq:mastereq_LEF} but with $\tilde{\alpha}$ replaced by $b$. However, the intensity dependence of $b = b(I,J)$, shown in Eq. \ref{eq:b_intdep} to be dominantly linear, results in a situation more akin to that detailed in Eq. \ref{eq:MEQ_satLEF_comov}. Importantly, this intensity dependence can create the necessary asymmetry between phase contributions arising from Kerr-like terms and those coming from direct phase modulation to enable coupling of the gain modulation into a phase modulation. For our simulations, we utilize a simplified model which accounts only for the net impact of this phase modulation, which we describe in the next section.

\subsubsection{Model simplifications}

Having discussed the master equation, including the LEF, as well as its relation to the Bloch gain, and explored how it can result in gain modulating coupling into phase, we now turn to deriving the exact model used in this work. We argue that the qualitative similarity between Eqs. \ref{eq:mastereq_Bloch} and \ref{eq:MEQ_satLEF_comov} justifies the use of Eq. \ref{eq:MEQ_satLEF_comov} as a starting point.

We proceed by looking at three groups of terms on the right-hand side, those containing only $A$ and polynomial orders of $A$, those containing $\frac{\partial A}{\partial t}$ and those containing $\frac{\partial^{2} A}{\partial t^{2}}$. For the terms containing $A$ and and polynomial orders thereof, we find

\begin{equation}
    \text{RHS}_{A} = \frac{g_{0}\bigl(1+m(\zeta-\frac{1}{k_{p}^{(1)}}t,t)\bigr)}{2(1+|A|^{2}/A_\text{sat}^{2})}A + i\tilde{\alpha}_{0}\bigl(1+\frac{|A|^{2}}{\tilde{A}_\text{sat}^{2}}\bigr)\frac{g_{0}\bigl(1+m(\zeta-\frac{1}{k_{p}^{(1)}}t,t)\bigr)}{2(1+|A|^{2}/A_\text{sat}^{2})}A-\frac{\alpha_{c}}{2}A.
\end{equation}

Here, the first term describes the saturating gain, and we keep it in its entirety. The second term contains the dominant contributions to the phase. Here, consistent with the prior discussion, inclusion of the intensity-dependent LEF results in three main contributions to the phase. The first is the direct phase term, and the other two are Kerr-like terms. Under the present form of equation \ref{eq:MEQ_satLEF_comov}, these two Kerr-like terms partially cancel. We therefore take the direct phase modulation term, ignoring the constant part (which only contributes as a constant roundtrip phase shift), as dominating and drop the remaining terms. Finally, we have the loss term. Next, we look at terms containing $\frac{\partial A}{\partial t}$ and find they simplify to a single term,

\begin{equation}
    \text{RHS}_\frac{\partial A}{\partial t} = -\frac{g_{0}\bigl(1+m(\zeta-\frac{1}{k_{p}^{(1)}}t,t)\bigr)}{2(1+|A|^{2}/A_\text{sat}^{2})}T_{2}\frac{\partial A}{\partial t}.
\end{equation}

This term looks like a nonlinear group velocity term for the field. The dominant contribution could ultimately be accounted for as a slight shift in the co-moving reference frame under consideration and does not have a significant impact on the dynamics, so we choose to ignore it. Finally, we look at terms containing $\frac{\partial^{2} A}{\partial t^{2}}$ and find

\begin{multline}
    \text{RHS}_{\frac{\partial^{2} A}{\partial t^{2}}} = i\frac{k_{p}^{(2)}}{2}\frac{\partial^{2}A}{\partial t^{2}} +\frac{g_{0}\bigl(1+m(\zeta-\frac{1}{k_{p}^{(1)}}t,t)\bigr)}{2(1+|A|^{2}/A_\text{sat}^{2})}\frac{T_{2}^{2}}{1+\tilde{\alpha}^{2}}\frac{\partial^{2} A}{\partial t^{2}}\\-i\tilde{\alpha}\frac{g_{0}\bigl(1+m(\zeta-\frac{1}{k_{p}^{(1)}}t,t)\bigr)}{2(1+|A|^{2}/A_\text{sat}^{2})}\frac{T_{2}^{2}}{1+\tilde{\alpha}^{2}}\frac{\partial^{2} A}{\partial t^{2}}.
\end{multline}

The first term is the group velocity dispersion (GVD) term. Likewise, the third term looks like a nonlinear GVD term. As with the group velocity term, this term can be dominantly accounted for as a shift in the considered value of $k_{p}^{(2)}$, so we accordingly drop the additional term. The second term defines the gain curvature. Here, the modulation is not significant in comparison to the larger static contribution and, moreover, the key value is the value at saturation, which we take to be constant. Thus, we replace it with the term $\frac{D_{g}}{2}\frac{\partial^{2}A}{\partial t^{2}}$, with constant $D_{g}$. Notably, however, $D_{g} \propto g_{0}T_{2}^{2}$ and furthermore would be expected to decrease with increasing intracavity intensity as well as with increasing $\tilde{\alpha}$. Putting all of these terms together we arrive at the final master equation,

\begin{equation}\label{eq:MEQ_simp}
    \frac{\partial A}{\partial \zeta} = i\frac{k_{p}^{(2)}}{2}\frac{\partial^{2}A}{\partial t^{2}} + \frac{g_{0}(1 + m(\zeta-\frac{1}{k_{p}^{(1)}}t,t))}{2(1+|A|^{2}/A_\text{sat}^{2})}A + iM(\zeta-\frac{1}{k_{p}^{(1)}}t,t)A-\frac{D_{g}}{2}\frac{\partial^{2}A}{\partial t^{2}} - \frac{\alpha_{c}}{2}A,
\end{equation}

\noindent where $M = \frac{g_{0}\tilde{\alpha}_{0}}{2}m$. Equation \ref{eq:MEQ_simp} is equivalent to the typical model employed for studying quantum walk comb dynamics, as in ref.~\cite{Dikopoltsev2025theoryQW}. The crucial final step for accurate simulation of the THz QCL device is accurate determination of $M(\zeta-\frac{1}{k_{p}^{(1)}}t,t)$, which we discuss in the next section.

\subsubsection{Microwave propagation}

Modeling of the laser dynamics requires active modeling of the microwave propagation in the device. Here, we extend the framework presented in~\cite{Schreiber2025_model} to incorporate the changing conductance and capacitance of the QCL gain region as a function of the applied voltage. We consider a ring resonator transmission line with $C_\text{tot}'(z,t)$, $G_\text{tot}'(z,t)$, $L'$, and $R'$ are the distributed capacitance, conductance, inductance, and resistance, as depicted in Fig. XX. Letting $u(z,t)$ and $i(z,t)$ be the voltage and capacitance across the line, we find the following set of coupled equations describing their evolution,

\begin{subequations}
    \begin{equation}\label{eq:TL_i}
        \frac{\partial i}{\partial z} = -C_\text{tot}'\frac{\partial u}{\partial t} - G_\text{tot}'u,
    \end{equation}
    \begin{equation}\label{eq:TL_u}
        \frac{\partial u}{\partial z} = -L'\frac{\partial i}{\partial t} - R'i.
    \end{equation}
\end{subequations}

Here, the active region contributes to both the capacitance and conductance, and both are taken to vary locally with the applied electric field. In our model, we consider a first-order approximation in which the quantities are allowed to vary linearly with the applied voltage,

\begin{subequations}
    \begin{equation}\label{eq:udep_C}
        C'_\text{tot}(z,t) = C' + \Delta C'u(z,t),
    \end{equation}
    \begin{equation}\label{eq:udep_G}
        G'_\text{tot}(z,t) = G' + \Delta G'u(z,t).
    \end{equation}
\end{subequations}

Differentiating Eq. \ref{eq:TL_i} with respect to $t$ and Eq. \ref{eq:TL_u} with respect to $z$ allows their combination into a single equation which, plugging in Eqs. \ref{eq:udep_C} and \ref{eq:udep_G}, may be written as

\begin{multline}\label{eq:NLTL_Full}
    \frac{\partial^{2} u}{\partial z^{2}} -L'C'\frac{\partial^{2} u}{\partial t^{2}} - (R'C'+L'G')\frac{\partial u}{\partial t}-R'G'u \\ = L'\Delta C'\biggl(u\frac{\partial^{2} u}{\partial t^{2}}+\bigl(\frac{\partial u}{\partial t}\bigr)^{2}\biggr) + R'\Delta C'u\frac{\partial u}{\partial t} + 2L'\Delta G'u\frac{\partial u}{\partial t} + R'\Delta G'u^{2}.
\end{multline}

Equation \ref{eq:NLTL_Full} represents the main result of this section. The left-hand side as written is a linear wave equation which describes the full lossy microwave propagation in the cavity. The right-hand side, meanwhile, represents the nonlinear contributions to the microwave propagation. On balance, the right-hand side provides a phase-sensitive quadratic nonlinearity for the microwave, which supports three-wave-mixing processes. We note here that the conclusions of this section are not restricted to the exact form of the nonlinearity presented but are general for any such phase-sensitive quadratic nonlinearity.

To proceed in our analysis, we may first consider the linear modes of the system, which may be computed by solving the wave equation

\begin{equation}\label{eq:lin_TL}
    \frac{\partial^{2} u_\text{lin}}{\partial z^{2}} -L'C'\frac{\partial^{2} u_\text{lin}}{\partial t^{2}} - (R'C'+L'G')\frac{\partial u_\text{lin}}{\partial t}-R'G'u_\text{lin} = 0.
\end{equation}

In frequency domain, this may be rewritten as

\begin{equation}\label{eq:lin_TL_FD}
    \frac{\partial^{2}\tilde{u}_\text{lin}}{\partial z^{2}} + \beta(\omega)^{2}\tilde{u}_\text{lin} = 0,
\end{equation}

\noindent where $\tilde{u}_\text{lin}(z,\omega) = \mathcal{F}\{u_\text{lin}(z,t)\}$ and the complex propagation constant $\beta(\omega)$ solves $\beta(\omega)^{2} = L'C'\omega^{2} - i\omega(R'C' + L'G')-R'G'$. With appropriate boundary conditions, solutions may readily be found. In our cavity, we assume a localized injection into the transmission line at $z = 0$ from a source with negligible source impedance. If we assume the cavity spans from $0$ to $L$ and consider a cosinusoidal injection, the appropriate boundary conditions are thus $u(0,t) = u(L,t) = u_\text{mod}\cos{(\omega_\text{mod}t)}$ and, by symmetry, $\frac{\partial u}{\partial z}|_{z=\frac{L}{2}}=0$. Then, the linear solutions of the voltage distribution in the cavity are given by

\begin{equation}\label{eq:lin_soln}
    u_\text{lin}(z,t) = u_\text{mod}\Re\bigg\{\frac{\cos{\bigl(\beta(\omega_\text{mod})(z-\frac{L}{2})\bigr)}}{\cos{\bigl(\beta(\omega_\text{mod})\frac{L}{2}\bigr)}}\bigg\}.
\end{equation}

Full solutions to Eq. \ref{eq:NLTL_Full} may be found via several different approaches. In our simulations, we use finite differencing methods to compute the evolution of the voltage along the transmission line. Specifically, the voltage at the $i^\text{th}$ spatial location for the $j+1^\text{th}$ time step is computed by solving

\begin{multline}\label{eq:FDTD_NLTL}
    \frac{-L'\Delta C'}{4\Delta t^{2}}(u_{j+1}^{i})^2 \\ + \biggl(\frac{L'\Delta C'}{2\Delta t^2}u_{j+1}^{i} - \frac{(L'\Delta G' + R'\Delta C')}{2\Delta t}u_{j}^{i} - \frac{L'\Delta C'}{\Delta t^2}u_{j}^{i} - \frac{(R'C' + L'G')}{2\Delta t} - \frac{L'C'}{\Delta t^2}\biggr)u_{j+1}^{i} \\ + \frac{u_{j}^{i-1} - 2u_{j}^{i} + u_{j}^{i+1}}{\Delta z^2} + L'C'\frac{2u_{j}^{i}-u_{j-1}^{i}}{\Delta t^2} + (R'C' + L'G')\frac{u_{j-1}^{i}}{2\Delta t} - R'G'u_{j}^{i} \\ + L'\Delta C'\frac{2u_{j}^{i}-u_{j-1}^{i}}{\Delta t^2}u_{j}^{i} - L'\Delta C'\frac{(u_{j-1}^{i})^2}{4\Delta t^2} + (L'\Delta G' + R'\Delta C')\frac{u_{j}^{i}u_{j-1}^{i}}{2\Delta t} - R'\Delta G'(u_{j}^{i})^2 \\ = 0
\end{multline}

\noindent with the above-given boundary conditions, where $\Delta t$ and $\Delta z$ are the step sizes of the discrete temporal and spatial grids used in our simulation. After solving for $u(z,t)$ in this way, the current density $J(z,t)$ is readily found through the relations $i'_\text{QCL}(z,t) = G_\text{tot}^{'}(z,t)u(z,t)$ and $J(z,t) = i'_\text{QCL}(z,t)/w$, where $w$ is the width of the active region.

In our simulations, we assume that $J_\text{DC}$ is only weakly impacted through nonlinear interaction with the high-frequency components and does not vary significantly in space, consistent with the linear solution given by Eq. \ref{eq:lin_soln} at DC. Moreover, the nonlinear transmission line model we consider already assumes an above-threshold QCL, so the values of $R'$, $L'$, $C'$, and $G'$ already account for this DC drive. Thus, only $J_\text{mod}(z,t)$ is computed using Eq. \ref{eq:NLTL_Full}. Finally, since $m(z,t) \propto J_\text{mod}(z,t)$, we ultimately relate $m(z,t) = \kappa i'_\text{QCL}(z,t)$ with $\kappa$ the proportionality constant.

\subsubsection{Modeling of subharmonic comb formation}

To summarize the above, modeling of the subharmonic comb formation in our system is achieved through simultaneous solution of Eqs. \ref{eq:MEQ_simp} and \ref{eq:NLTL_Full}. For each step in time, Eq. \ref{eq:NLTL_Full} enables calculation of the instantaneous microwave profile in the cavity using finite differences. The result enables solution of Eq. \ref{eq:MEQ_simp} for the instantaneous optical field profile, $A(z,t)$, using a Fourier split-step approach.

The parameters used in our simulations are given by Table \ref{tab:sim_params}.

\begin{table}[h!]
\centering
\begin{tabular}{||c c c||}
    \hline Parameter & Value & Units\\
    \hline\hline $L_\text{cav}$ & $\pi\times$1.6e-3 & m \\
    \hline $D_{g}L_\text{cav}$ & 5.25e-28 & s$^{2}$ \\
    \hline $k_{p}^{(2)}L_\text{cav}$ & 5e-25 & s$^{2}$ \\
    \hline $\alpha_{c}L_\text{cav}$ & 3.5 & \\
    \hline $g_{0}L_\text{cav}$ & 5.25 & \\
    \hline $\tilde{\alpha}_{0}$ & 2 &  \\
    \hline $A_\text{sat}^{2}$ & 1e-3 & W \\
    \hline FSR & 15.89e9 & Hz \\
    \hline $\kappa$ & 5.75e-6 & A$^{-1}$~m \\
    \hline $R'$ & 568 & $\Omega~\text{m}^{-1}$ \\
    \hline $L'$ & 9.86e-8 & H~m$^{-1}$ \\
    \hline $C'$ & 6e-10 & F~m$^{-1}$ \\
    \hline $G'$ & 38 & S~m$^{-1}$ \\
    \hline $\Delta C'$ & -1.2e-11 & F~m$^{-1}$~V$^{-1}$ \\
    \hline $\Delta G'$ & -3.8 & S~m$^{-1}$~V$^{-1}$ \\
    \hline
\end{tabular}
\caption{Simulation parameters.}
\label{tab:sim_params}
\end{table}

\begin{figure}[t!]
\centering
\includegraphics[width=0.95\linewidth]{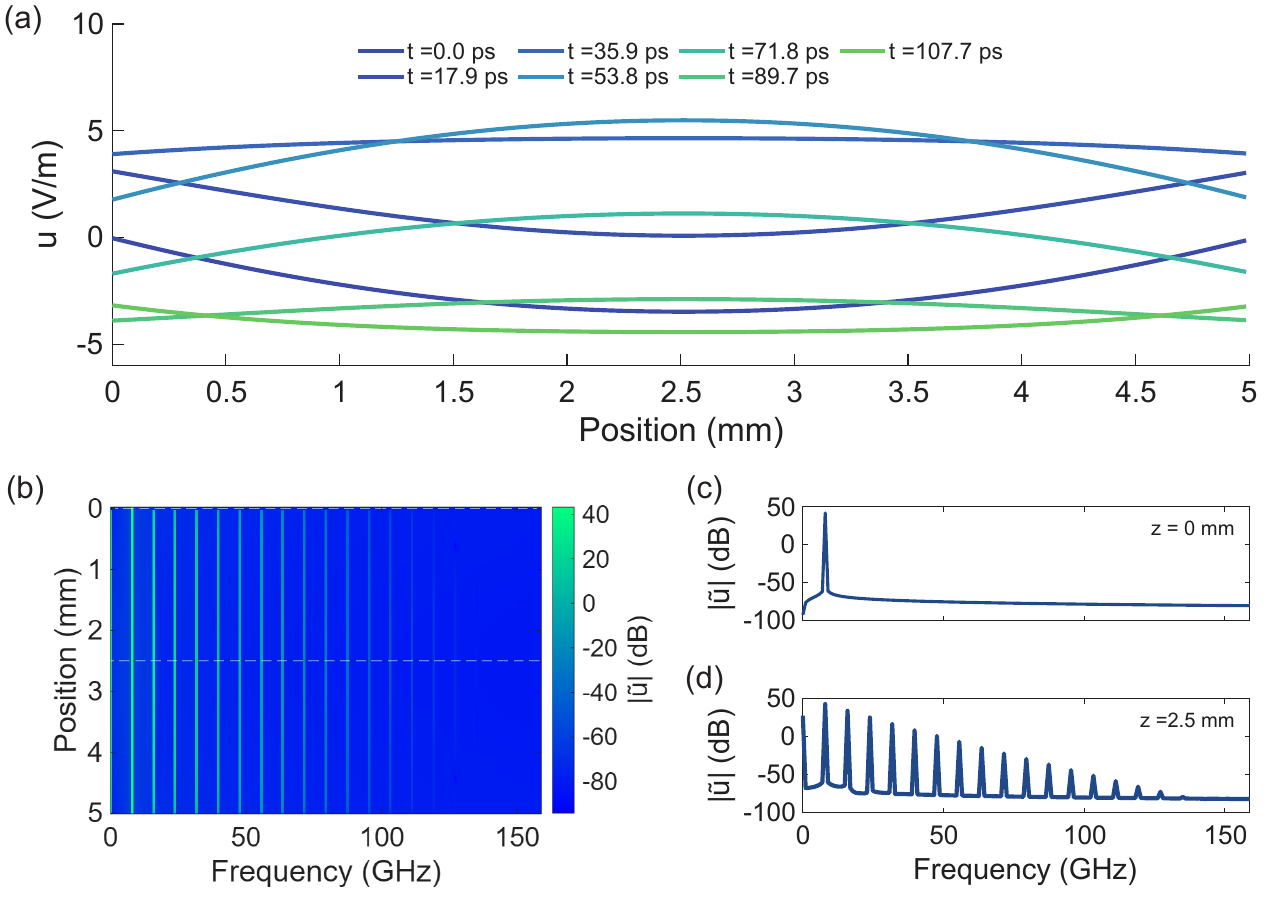}
\caption{Nonlinear microwave propagation. (a) Temporal evolution of $u$ as a function of cavity position. Here, modulation at FSR/2 is considered over a time window of $2T_\text{RT}$. (b) Spectral amplitude of the modulation voltage as a function of position in the cavity. (c) A pure tone at $f_\text{rep}/2$ is injected at $z = 0$ mm, and (d) the nonlinear propagation leads to the generation of many harmonics.}
\label{sfig:theory1}
\end{figure}

Figure \ref{sfig:theory1} illustrates the nonlinear microwave propagation in the case of $f_\text{mod} = \text{FSR/2}$. The microwave voltage profile, $u(z,t)$, for several instances of time within one period of the modulation is shown in Figure \ref{sfig:theory1}a. The input, $u(0,t)$ is taken to be a sinusoid with amplitude $u_\text{mod} = 4$ V. In Fig. \ref{sfig:theory1}b, we plot $|\tilde{u}|$, where $\tilde{u}(z,t) = \mathcal{F}\{u(z,t)\}$, revealing the consequence of the nonlinear propagation on the frequency content of $u$. While at $z = 0$, $u$ consists of a single frequency component at the drive (Fig. 2c), the nonlinear propagation leads to the generation of many high harmonics (d). The strong component at the FSR is dominantly responsible for the QW comb formation under subharmonic injection. In our simulations, we use $u_\text{mod} = 1.1$ V for the fundamental QW at $f_\text{mod} = \text{FSR}$, $u_\text{mod} = 4$ V for $f_\text{mod} = \text{FSR}/2$, $u_\text{mod} = 7$ V for $f_\text{mod} = \text{FSR}/3$, and $u_\text{mod} = 9$ V for $f_\text{mod} = \text{FSR}/4$. While in our experiment, similar RF drive powers at the source are used in all cases, this scaling of the on-chip modulation strength is consistent with higher losses expected in the cabling and printed circuit board used to deliver the RF to the QCL device at higher modulation frequencies.

\begin{figure}[t!]
\centering
\includegraphics[width=0.95\linewidth]{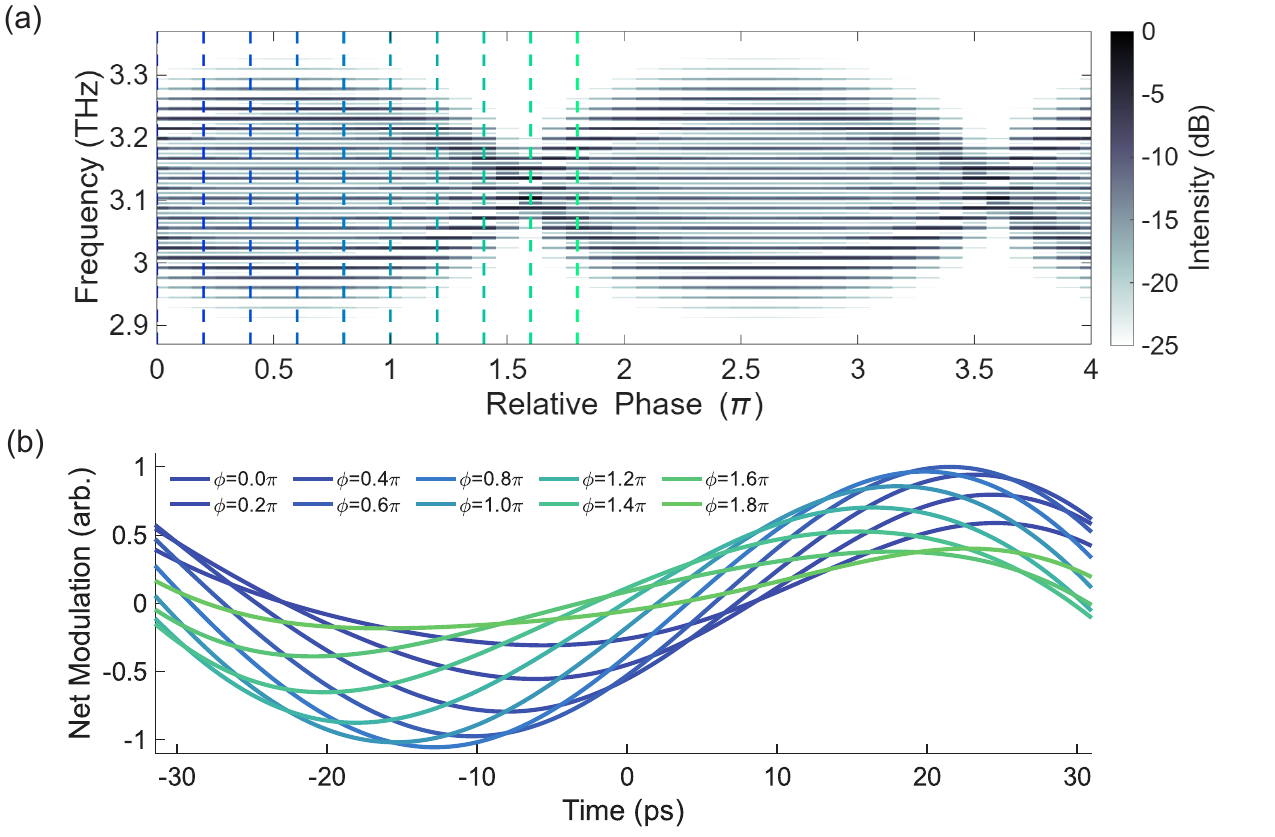}
\caption{Two-tone control of comb profile. (a) Simulated THz comb output under two-tone injection at FSR and FSR/2, as in Fig. 5b. (b) Normalized profile of the net phase modulation over one roundtrip for different relative phases, $\phi$, at positions indicated in (a).}
\label{sfig:theory2}
\end{figure}

Next, in Fig. \ref{sfig:theory2}, we further analyze the case of two-tone modulation. Here, $u(0,t) = u_\text{mod,1}\sin{(\omega_{1}t + \phi)} + u_\text{mod,2}\sin{(\omega_{2}t)}$. Figure \ref{sfig:theory2} is a repeat of the simulation of Fig. 5b in the main text, where $f_\text{mod,1}$ = FSR and $f_\text{mod,2}$ = FSR/2 while $u_\text{mod,1} = 0.6$ V and $u_\text{mod,2} = 4$ V. We refer to $\phi$ as the relative phase. In the case of $f_\text{mod,1}$ = FSR and $f_\text{mod,2}$ = FSR/3, we use $u_\text{mod,1} = 0.6$ V and $u_\text{mod,2} = 7$ V, while in the case of $f_\text{mod,1}$ = 2FSR/3 and $f_\text{mod,2}$ = FSR/3, we also use $u_\text{mod,1} = 0.6$ V and $u_\text{mod,2} = 7$ V.

To understand the behavior of the two-tone modulation, we isolate the phase modulation term in Eq. \ref{eq:MEQ_simp}, $\frac{\partial A}{\partial \zeta} \propto iM(\zeta - \frac{1}{k_{p}^{(1)}}t,t)A$, which we identify as a convolution over the spatial coordinate $\zeta$. From this, the phase accumulated over one roundtrip by a given point $t$ on the wave is $\Delta\phi_\text{net}^\text{RT}(t) \approx \int_{0}^{L_\text{cav}}M(\zeta-\frac{1}{k_{p}^{(1)}}t,t)d\zeta$. Crucially, this leads to a phase-matching condition, that only modulation components which have a phase velocity equal to the group velocity of the THz field couple efficiently.

Thus, the RF nonlinearity can play two roles in the modulation of the THz field. Firstly, the nonlinear mixing of the microwaves can change their relative strength in the cavity. Secondarily, the nonlinearity can lead to an effective propagation constant, $\beta_\text{RF}^\text{NL}$, which differs from the linear propagation constant. When this effective propagation constant is made to better satisfy the phase-matching condition, more efficient modulation is observed. In the case of the two-tone modulation, we find this to be the dominant effect which leads to the observe comb shaping. These effects are jointly captured in $\Delta\phi_\text{net}^\text{RT}$, the AC component of which is plotted in Fig. \ref{sfig:theory2}b for 10 different values of the relative phase, corresponding to the vertical dashed lines of Fig. \ref{sfig:theory2}a. Here, we see that at $\phi = 0.6\pi$, where the comb bandwidth is strongest, the peak-to-peak strength of the net modulation is maximized, while for $\phi = 1.6\pi$, where the comb bandwidth is smallest, the net modulation is minimized. Combined, these results indicate how the nonlinear microwave propagation can be leveraged to influence the comb behaviors.

\newpage
\printbibliography
